\documentclass[review]{elsarticle}

\usepackage{textcomp}
\usepackage{lineno}
\usepackage{amssymb}
\usepackage{mathrsfs}
\usepackage[colorlinks,citecolor=blue,urlcolor=blue,linkcolor=blue]{hyperref}
\usepackage{color}

\modulolinenumbers[1]

\journal{Nuclear Instruments and Methods A}

\begin{document}

\begin{frontmatter}

\title{PICOSEC: Charged particle timing at sub-25 picosecond precision with a Micromegas based detector}

\author[CERN]{J.~Bortfeldt}
\author[CERN]{F.~Brunbauer}
\author[CERN]{C.~David}
\author[CEA]{D.~Desforge}
\author[NSCR]{G.~Fanourakis}
\author[CERN]{J.~Franchi}
\author[LIPP]{M.~Gallinaro}
\author[CEA]{I.~Giomataris}
\author[IGFAE]{D.~Gonz\'alez-D\'iaz}
\author[CEA]{T.~Gustavsson}
\author[CEA]{C.~Guyot}
\author[CEA]{F.J.~Iguaz\corref{mycorrespondingauthor}}
\ead{iguaz@cea.fr}
\author[CEA]{M.~Kebbiri}
\author[CEA]{P.~Legou}
\author[USTC]{J.~Liu}
\author[CERN]{M.~Lupberger}
\author[CEA]{O.~Maillard}
\author[AUTH]{I.~Manthos}
\author[CERN]{H.~M\"uller}
\author[AUTH]{V.~Niaouris}
\author[CERN]{E.~Oliveri}
\author[CEA]{T.~Papaevangelou}
\author[AUTH]{K.~Paraschou}
\author[CEA]{M.~Pomorski}
\author[USTC]{B.~Qi}
\author[CERN]{F.~Resnati}
\author[CERN]{L.~Ropelewski}
\author[AUTH]{D.~Sampsonidis}
\author[CERN]{T.~Schneider}
\author[CEA]{P.~Schwemling}
\author[CERN]{L.~Sohl}
\author[CERN]{M.~van~Stenis}
\author[CERN]{P.~Thuiner}
\author[NTUA]{Y.~Tsipolitis}
\author[AUTH]{S.E.~Tzamarias}
\author[RD51]{R.~Veenhof\fnref{Veenhof}}
\author[USTC]{X.~Wang}
\author[CERN]{S.~White\fnref{Virginia}}
\author[USTC]{Z.~Zhang}
\author[USTC]{Y.~Zhou}

\address[CEA]{IRFU, CEA, Universit\'e Paris-Saclay, F-91191 Gif-sur-Yvette, France}
\address[CERN]{European Organization for Nuclear Research (CERN), CH-1211 Geneve 23, Switzerland}
\address[USTC]{State Key Laboratory of Particle Detection and Electronics, University of Science and Technology of China, Hefei 230026, China}
\address[AUTH]{Department of Physics, Aristotle University of Thessaloniki, Thessaloniki, Greece}
\address[NSCR]{Institute of Nuclear Physics, NCRS Demokritos, 15310 Aghia Paraskevi, Athens, Greece}
\address[NTUA]{National Technical University of Athens, Athens, Greece}
\address[LIPP]{Laborat\'orio de Instrumenta\c{c}\~ao e F\'isica Experimental de Part\'iculas, Lisbon, Portugal}
\address[RD51]{RD51 collaboration, European Organization for Nuclear Research (CERN), CH-1211 Geneve 23, Switzerland}
\address[IGFAE]{Instituto Galego de F\'isica de Altas Enerx\'ias (IGFAE), Universidade de Santiago de Compostela, Spain}

\cortext[mycorrespondingauthor]{Corresponding author}
\fntext[Veenhof]{Also at National Research Nuclear University MEPhI, Kashirskoe Highway 31, Moscow, Russia;
and Department of Physics, Uluda\u{g} University, 16059 Bursa, Turkey.}
\fntext[Virginia]{Also at University of Virginia}

\begin{abstract}
The prospect of pileup induced backgrounds at the High Luminosity LHC (HL-LHC)
has stimulated intense interest in developing technologies for charged particle detection with accurate timing at high rates.
The required accuracy follows directly from the nominal interaction distribution
within a bunch crossing ($\sigma_z\sim5$\,cm, $\sigma_t\sim170$\,ps).
A time resolution of the order of 20-30\,ps would lead to significant reduction of these backgrounds.
With this goal, we present a new detection concept called PICOSEC,
which is based on a ``two-stage'' Micromegas detector coupled to a Cherenkov radiator and equipped with a photocathode.
First results obtained with this new detector yield a time resolution of 24\,ps for 150\,GeV muons,
and 76\,ps for single photoelectrons.
\end{abstract}


\end{frontmatter}


\section{Introduction}
\label{sec:Introduction}
The prospect of pileup induced backgrounds at the High Luminosity LHC (HL-LHC)
has stimulated interest in developing technology for charged particle timing at high rates~\cite{White:2013taa}.
Since the hermetic timing approach
(where a large fraction of tracks are used to time interaction vertices)
requires a large area coverage, it is natural to investigate
both MicroPattern Gas and Silicon structures as candidate detector technologies
to address this approach.
However, since the necessary time resolution for pileup mitigation is of the order of 20-30 picoseconds (ps),
both technologies require significant modification to reach the desired performance.

Photodetectors and charged particle detectors with time resolutions in the sub-nanosecond regime
continue to have an impact in both High Energy physics and medical imaging.
In High Energy physics the most widespread application is for particle identification, 
wherein the mass of particles of known momentum 
is measured with the time-of-flight (TOF) particle identification technique.
The current state-of-the-art technology has recently been reviewed in Ref.~\cite{Vavra:2017jv}.
Existing collider experiments (e.g. ALICE) now employ large scale TOF systems with performance at the sub-100\,ps level
but several promising new technologies have demonstrated $\sim10$\,ps performance or better
(for example the Microchannel PMT we use as a Cherenkov detector during the test beam measurements discussed below).

An early incentive for the development of MicroPattern Detectors
(see e.g.\,\cite{Heijne:1979ub}) was the promise of faster response and improved timing
precision - a consequence of the more rapid signal collection
and lower sensor capacitance.
Except for early work at CERN (e.g. 1970-80s in the case of~\cite{Heijne:1979ub}),
however, the emphasis in MicroPattern Silicon detectors rapidly moved
to spatial resolution rather than temporal precision.
Similarly, there has been little emphasis, in the 20 years since the GEM~\cite{Sauli:1997qp}
and Micromegas~\cite{Giomataris:1995fq}
MicroPattern Gas Detectors (MPGD) were introduced, on exploiting their potential for fast timing.
Nevertheless, in one of the original Micromegas papers~\cite{Giomataris:1998rc}, it was shown 
that sub-nanosecond time jitter could be obtained for single electrons photo-produced at the cathode surface.
This is the approach that will be followed in the current paper, i.e.
the timing attributes of Micromegas are used in a photodetector.

In 2015, our collaboration proposed a structure detecting extreme ultraviolet (UV) Cherenkov light produced
in a MgF$_2$ crystal coupled to semitransparent CsI photocathode
and a two-stage Micromegas amplifying structure with electron amplification in both stages.
This detetector has several advantages compared to a single-stage structure:
higher gain, a reduced ion-backflow and a better separation between the electron-peak and the ion-tail of the signal.
In Ref.~\cite{Papaevangelou:2016knm}, we reported encouraging results of single photoelectron timing resolution
using fast laser pulses,  obtained with a chamber operated in \textit{sealed mode}.
Subsequently, we improved the chamber integrity and the detector grounding in order to guarantee stable operation.
For the results presented here we concentrate on a single pixel PICOSEC chamber (1\,cm diameter),
and use a bulk Micromegas amplification structure with a woven stainless steel mesh~\cite{Giomataris:2004aa}.
Beam tests using 150 GeV muons at CERN were carried out
using several Micromegas detectors with various photocathode materials,
gases, different discharge protection schemes, and read-out elements.

In this paper we present the main results obtained,
demonstrating the capability of our detector to reach time resolution of tens of ps, 
an improvement by two orders of magnitude compared to the standard MPGD detector performance.
Detailed analysis methods and simulations were developed to better understand
single electron detector response and time resolution.

The manuscript is divided as follows: the PICOSEC detection concept is presented in Sec.~\ref{sec:Concept},
while the technical description of the first prototype is given in Sec.~\ref{sec:Prototype}.
The experimental setups used to measure the time resolution of
single photoelectrons with the laser and 150\,GeV muons in the beam
are detailed in Sec.~\ref{sec:LaserSetup} and Sec.~\ref{sec:BeamSetup}, respectively.
After a description of the individual waveform analysis in Sec.~\ref{sec:PulseAnalysis},
the laser tests results are presented in Sec.~\ref{sec:ResultsLaser},
while those from beam tests are summarized in Sec.~\ref{sec:ResultsBeam}.
The conclusions in Sec.~\ref{sec:Conclusions} complete this paper.

\section{The PICOSEC detection concept and the experimental setup}
\label{sec:Concept}
\begin{figure}[htb!]
\centering
\includegraphics[width=0.99\textwidth]{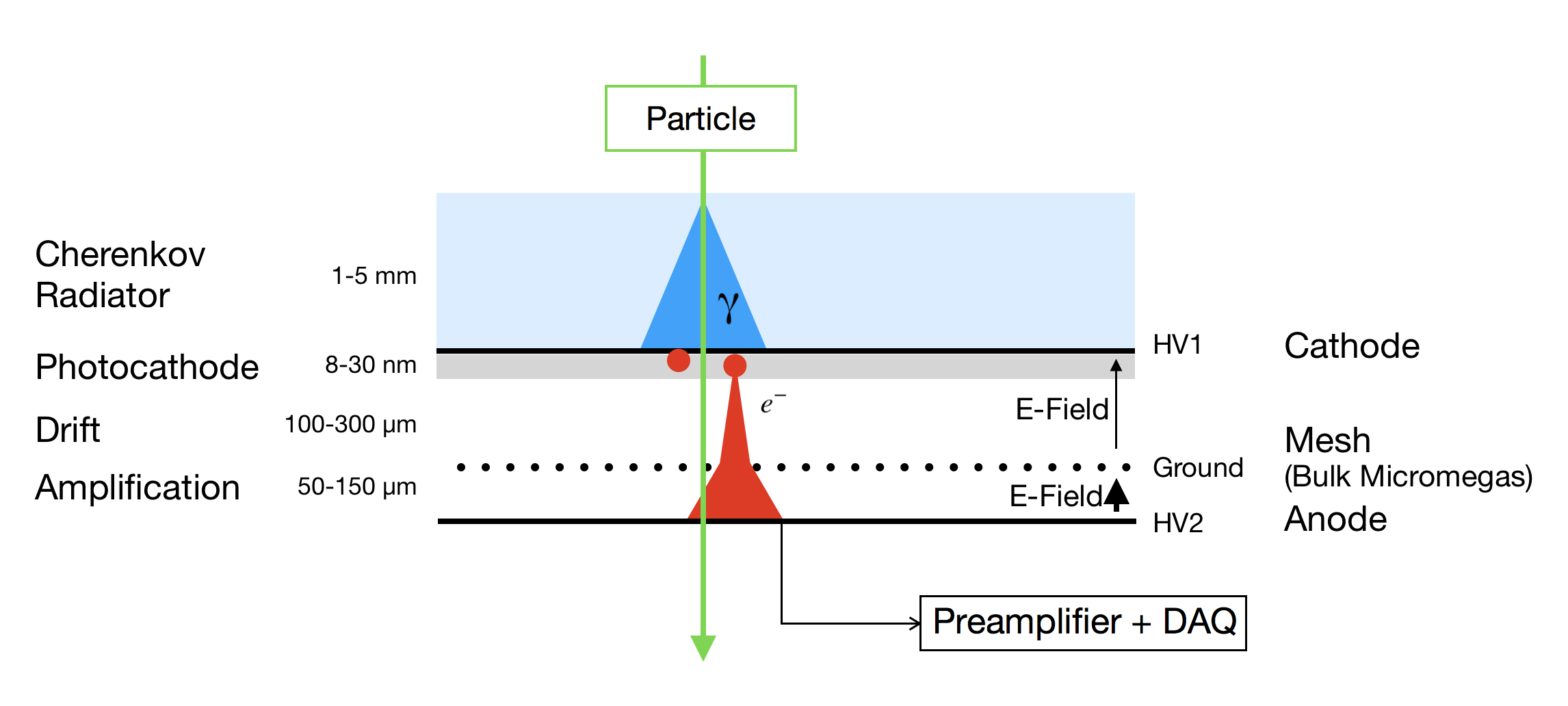}
\caption{The PICOSEC detection concept. The passage of a charged particle through the Cherenkov radiator
produces UV photons, which are then absorbed at the photocathode and partially converted into electrons.
These electrons are subsequently preamplified and then amplified in the two high-field drift stages,
and induce a signal which is measured between the anode and the mesh.}
\label{fig:DetectionConcept}
\end{figure}

The detection concept presented here consists of a ``two-stage'' Micromegas detector
coupled to a front window that acts as Cherenkov radiator coated with a photocathode,
as shown in Fig.~\ref{fig:DetectionConcept}.
A MgF$_2$ crystal is a typical radiator with light transmission down to a wavelength of~115\,nm,
a typical photocathode is CsI~\cite{SEGUINOT1990133} with high quantum efficiency for photons below 200\,nm.
This configuration provides a large bandwidth for Cherenkov light production-detection in the extreme UV.
The drift region is very thin (100-300\,\textmu m),
which minimizes diffusion effects on the signal timing (of several ns in a typical MPGD-based drift region).
Due to the high electric field, photoelectrons also undergo pre-amplification in the drift region.

The readout is a bulk Micromegas~\cite{Giomataris:2004aa},
which consists of a woven mesh (18\,\textmu m-diameter wires) and an anode plane,
separated by a gap of 128\,\textmu m, mechanically defined by pillars.
This type of readout, operated in Neon- or CF$_4$-based gas mixtures, can reach gains of $10^5-10^6$,
high enough to detect single photoelectrons~\cite{Derre:1999wh}.
Moreover, the electron drift velocity is high enough for a fast charge collection:
ranging from 9 to 16\,cm/\textmu s for drift fields between 10 and 20\,kV/cm
in the ``COMPASS'' gas (80\%Ne + 10\%C$_2$H$_6$ + 10\%CF$_4$),
according to simulations with Magboltz~\cite{BIAGI1999234}.

In normal operation, a relativistic charged particle traversing the radiator produces UV photons,
which are simultaneously (RMS less than 10\,ps) converted into primary (photo)~electrons at the photocathode.
These primary electrons are preamplified in the drift region due to the high electric field ($\sim$20\,kV/cm);
then, they partially traverse the mesh ($\sim$25\% due to the field configuration),
and are finally amplified in the amplification gap, where a high electric field ($\sim$40\,kV/cm) is applied.

\begin{figure}[htb!]
\centering
\includegraphics[width=0.80\textwidth]{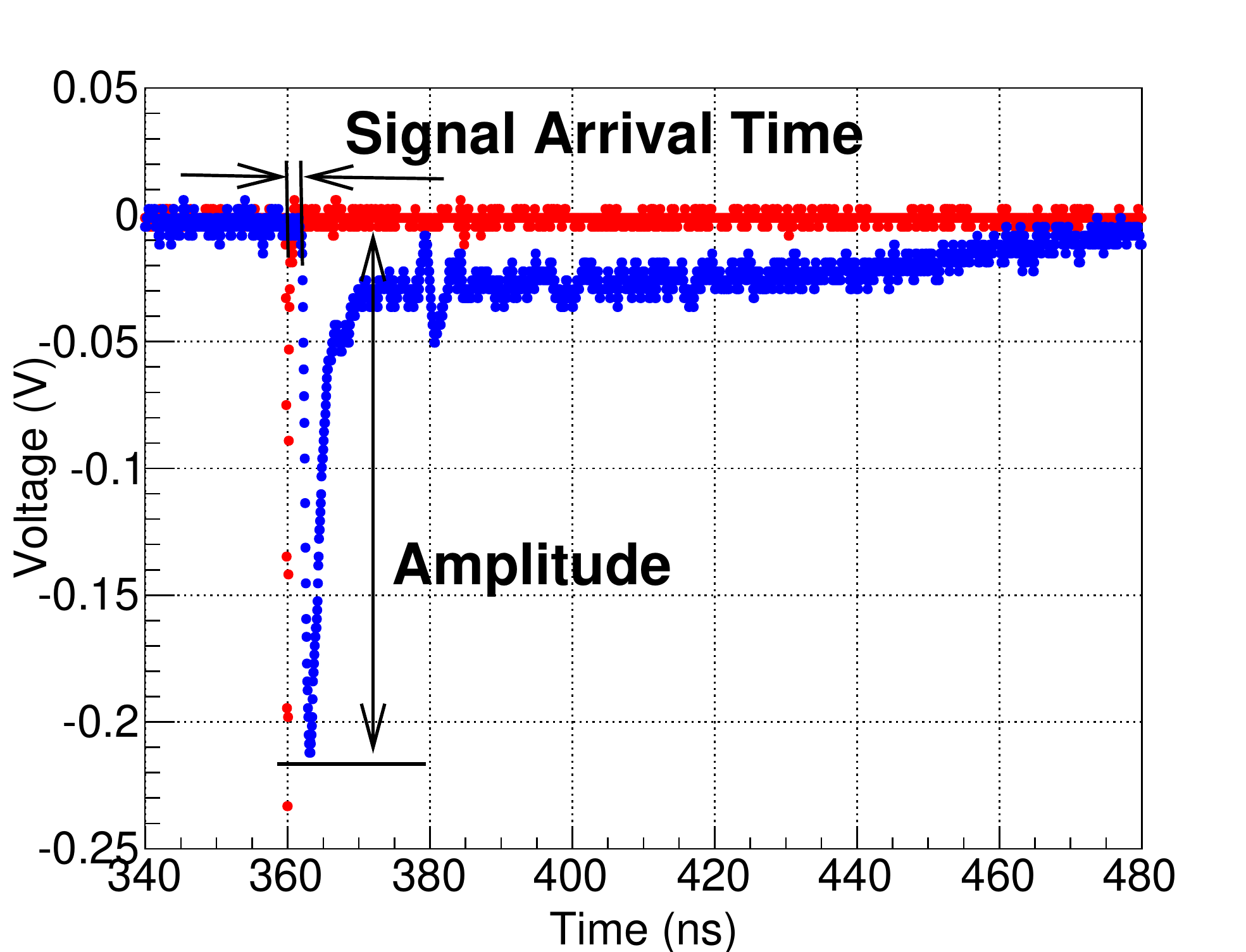}
\caption{An example of an induced signal from the PICOSEC detector generated by 150\,GeV muons (blue points),
recorded together with the timing reference of the microchannel plate MCP signal (red points) discussed in the text.
The PICOSEC signal contains a fast component produced by the electrons,
and a slower component generated by the ion drift.
The fast electron-peak amplitude and the Signal Arrival Time,
defined in the waveform analysis of Sec.~\ref{sec:PulseAnalysis}, are also shown.}
\label{fig:PicosecSignal}
\end{figure}

The arrival of the amplified electrons at the anode
produces a fast signal (with a risetime of $\sim$0.5\,ns) referred to as the electron-peak,
while the movement of the ions produced in the amplification gap generates a slower component - ion-tail ($\sim$100\,ns).
A typical waveform is shown in Fig.~\ref{fig:PicosecSignal}.
The maximum drift time of ions is below 630\,ns,
which is low enough not to affect the detector rate capability.

It should be noted that due to preamplification in the thin drift gap the relative contribution
to the overall signal of direct ionization produced by the traversing particle is negligible.
In the ``COMPASS gas'' and for the conditions described in Sec.~\ref{sec:BeamSetup},
relativistic muons create $\sim 21$~ion clusters/cm with few ionization electrons per cluster.
The probability to produce enough ionization charge that undergoes the same amplification (i.e. in the first $\sim$30 \,\textmu m)
as the typical 10 photoelectrons from the Cherenkov signal is only a few percent.

\subsection{Prototype description}
\label{sec:Prototype}
\begin{figure}[htb!]
\centering
\includegraphics[width=0.70\textwidth]{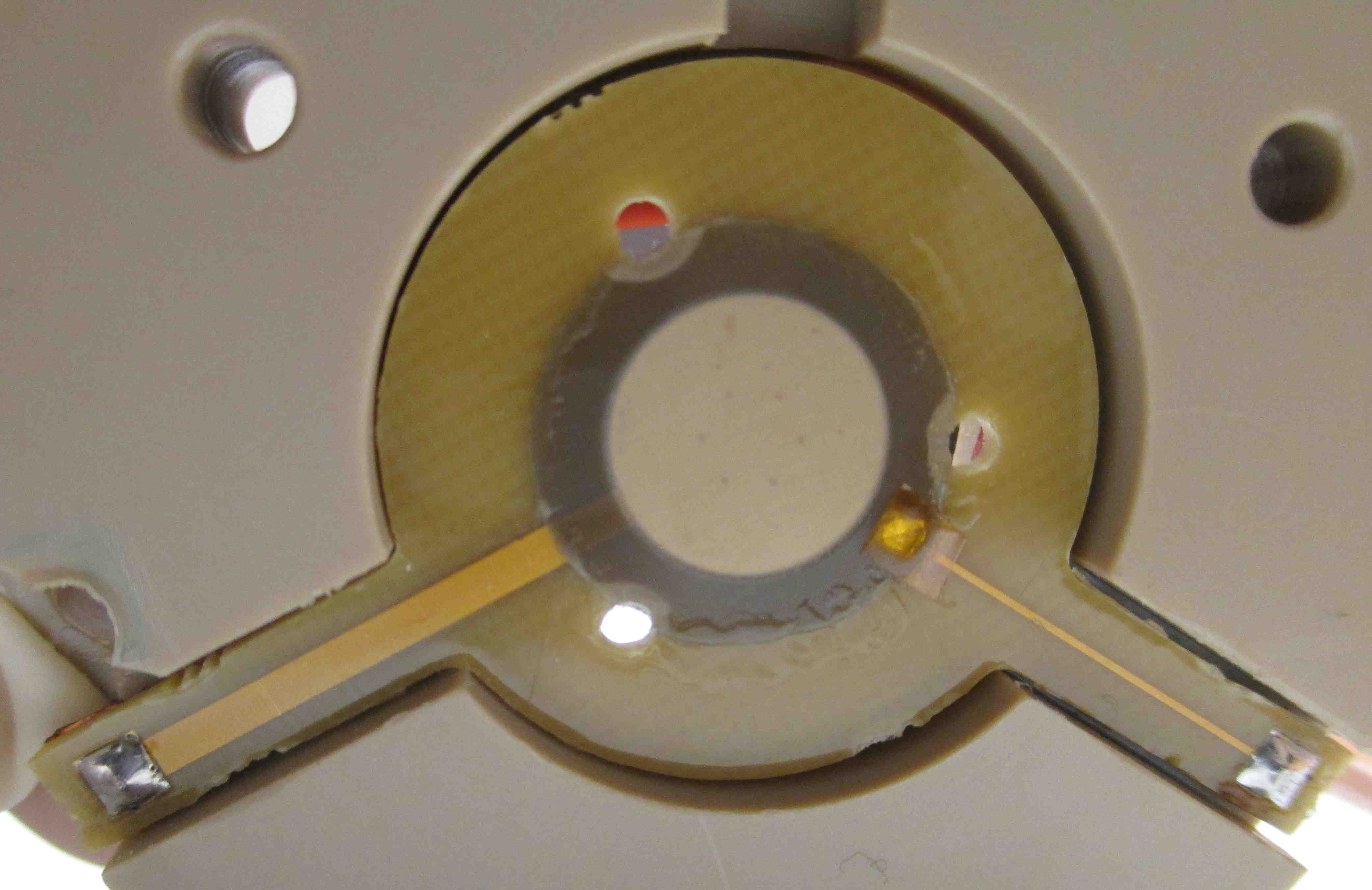}
\caption{Photograph of the readout structure of the bulk Micromegas detector in the first PICOSEC prototype.
Six pillars, arranged in a hexagonal pattern, support the mesh in the central region of the amplification gap.
The mesh and anode voltages are supplied by the two visible strip-lines
onto which two coaxial cables are soldered outside the active volume.
Their shielding is soldered to the solid copper ground layer on the lower side of the readout PCB.}
\label{fig:MMReadout}
\end{figure}

\begin{figure}[htb!]
\centering
\includegraphics[width=0.80\textwidth]{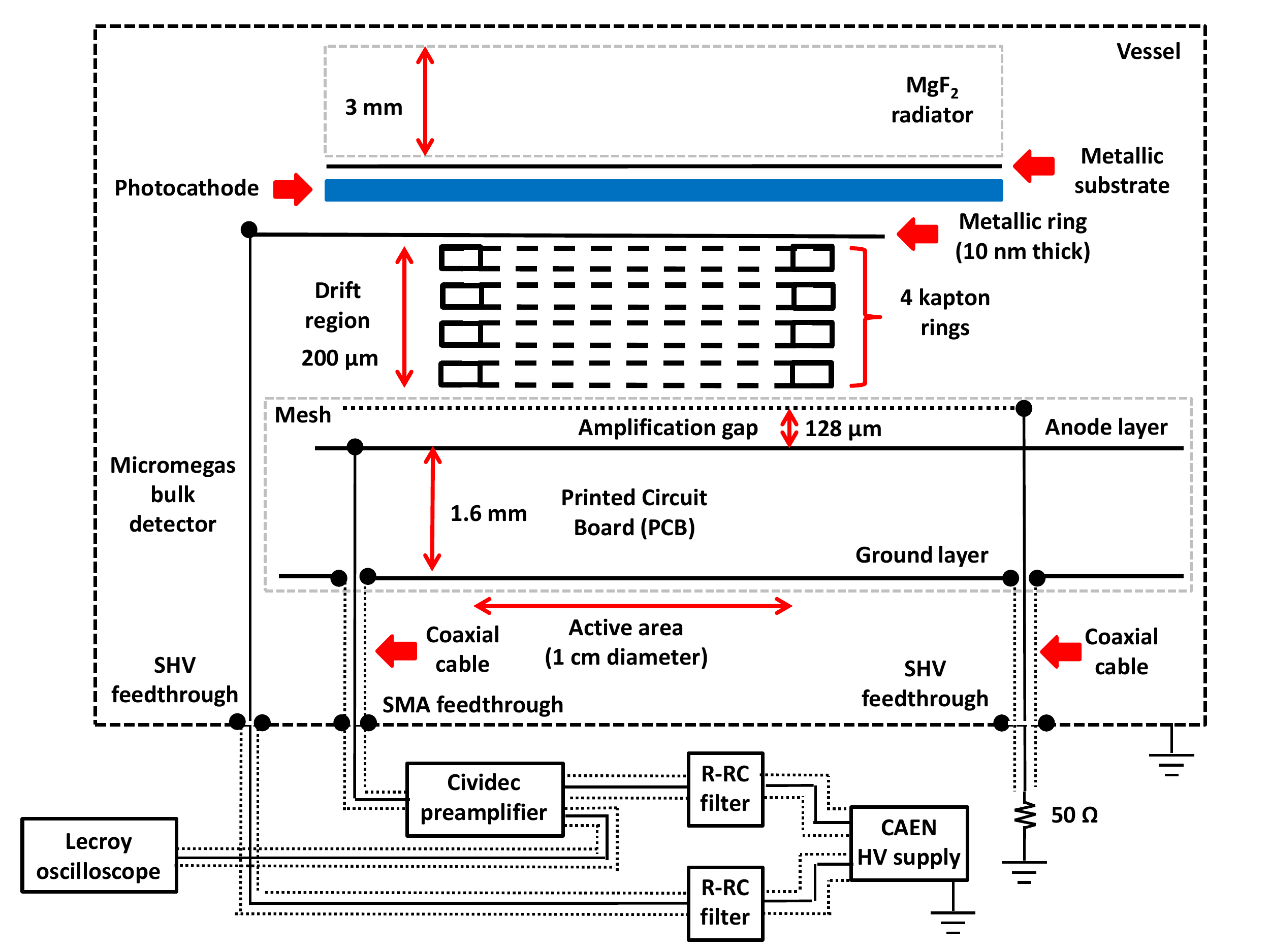}
\caption{Sketch of the first prototype of the PICOSEC detector, described in detail in the text.
The scale of some components is exaggerated for clarity.}
\label{fig:PrototypeSchema}
\end{figure}

A sketch of the first PICOSEC prototype is presented in Fig.~\ref{fig:PrototypeSchema}.
The readout (Fig.~\ref{fig:MMReadout}) is a bulk Micromegas detector
built on top of a 1.6\,mm thick Printed Circuit Board (PCB)
with a single anode (1\,cm diameter, 18\,\textmu m copper thickness) and
an amplification gap of 128\,\textmu m, i.e. the distance between the anode and the mesh wires.
The mesh used for the prototype is formed by 18\,\textmu m-diameter wires and has a 51\% optical transparency.
At the bottom part of the PCB, there is a 18\,\textmu m thick copper layer used as ground reference.
The amplification gap (between the mesh and the anode in Fig.~\ref{fig:DetectionConcept}) 
is defined by only six 200\,\textmu m-diameter pillars to minimize their influence in the two amplification stages.
The pillars are arranged in a circle and are fully contained in the sensitive area.
Four kapton rings, of 50\,\textmu m thickness each, are placed between the mesh and the crystal (i.e.~radiator)
to define the height of the drift gap of 200\,\textmu m.
During the laser tests, the crystal was made out of a 3\,mm thick MgF$_2$,
on which a 5.5\,nm chromium layer was deposited
serving as photocathode\footnote{A 1\,mm thick quartz crystal with a deposit
of 10\,nm Aluminum, or 100\,nm diamond was also tested with comparable results.}.
During beam tests, an additional 18\,nm-thick CsI layer was deposited over the chromium substrate
in order to increase the number of photoelectrons produced by charged particles.
In both configurations, a 10\,nm-thick metallic ring is placed on the crystal
to establish the potential of the photocathode.

The whole detector is installed inside a stainless steel chamber,
which is then filled with the ``COMPASS gas'' at 1\,bar absolute pressure.
Other gases, like 80\%CF$_4$+20\%C$_2$H$_6$ at 0.5\,bar absolute pressure, have also been used
but this article will focus on the results obtained with the ``COMPASS gas''.
The vessel has a transparent (quartz) entrance window to allow the passage of either UV~light or laser pulse; 
it has two gas valves for gas circulation, as well as a large vacuum port for evacuating the vessel.

Referring to Fig.~\ref{fig:PrototypeSchema}, cathode, anode and mesh elements
are electrically connected by SMA or SHV feedthroughs, as indicated.
The cathode is connected to one CAEN High Voltage Supply (HVS) channel,
the anode to a CIVIDEC preamplifier\footnote{\url{https://cividec.at/index.php?module=public.product&idProduct=34&scr=0}}
(2\,GHz, 40\,dB, a gain of 100), biased by a separate channel of the HVS,
and the mesh is connected to ground by a 50\,m long BNC cable,
terminated with a $50$\,Ohm resistor in order to avoid signal reflections.
Special attention was paid to proper grounding throughout the electronics design,
and referred to the ground layer of the Micromegas readout.
Each of the high voltage lines has a dedicated low-pass filter to suppress ripples from the HVS.

\subsection{Laser test: single photoelectron measurements}
\label{sec:LaserSetup}
\begin{figure}[htb!]
\centering
\includegraphics[width=0.90\textwidth]{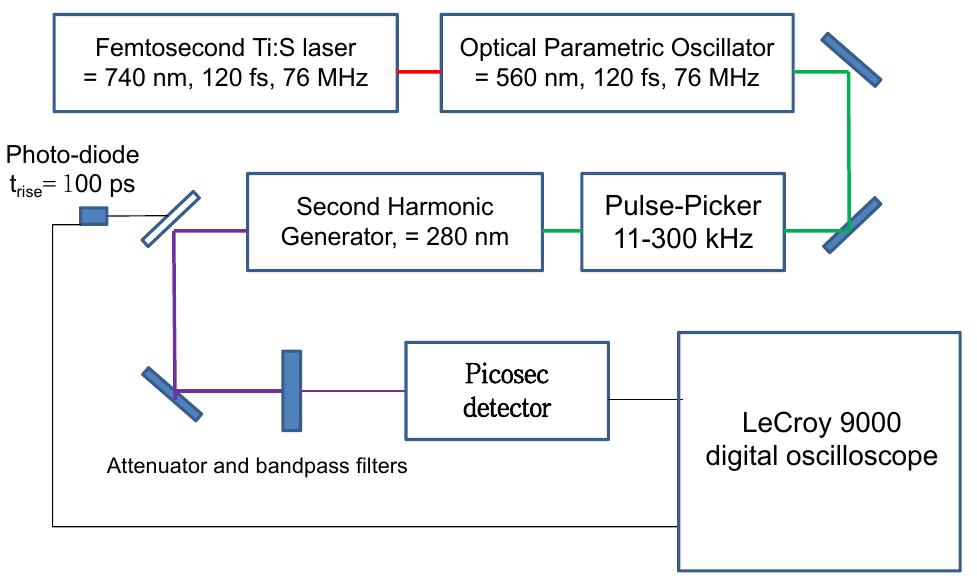}
\caption{Schematic of the experimental setup during the laser tests, described in detail in the text.}
\label{fig:LaserSchema}
\end{figure}

The time response of the PICOSEC detector for single photoelectrons
was measured at 
the Saclay Laser-matter Interaction Center (IRAMIS/SLIC, CEA).
The experimental setup (Fig.~\ref{fig:LaserSchema}) includes a femtosecond laser
with a pulse rate ranging from 9\,kHz to 4.7\,MHz at 267-288\,nm wavelength and a focal length of $\sim$1\,mm.
The laser beam is split into two equal parts, one arriving directly at the prototype and the other at a fast photodiode (PD0).
The PD0 signal serves as the reference time, simultaneously recorded with the PICOSEC detector signal during the measurements.
The high laser intensity at the PD0 and the fast risetime of this device
result in a reference time accuracy of approximately 13\,ps.
The intensity of the laser arriving at the detector is reduced by a series of light attenuators: 
electroformed fine nickel meshes (100-2000\,LPI) with optical transmission varying between 10\% and 25\%,
yielding attenuation factors of 4, 5, 10, and their combinations.
The Micromegas detector signal goes through
a CIVIDEC preamplifier before being digitized and registered together with the PD0
signal by a 2.5\,GHz oscilloscope at a rate of 20\,GSamples/s (i.e. one sample every 50\,ps).

The PICOSEC detector was operated with the ``COMPASS gas'' at 1\,bar absolute pressure.
The anode voltage (HV2 in Fig.~\ref{fig:DetectionConcept}) was scanned between 450\,V and 525\,V in steps of 25\,V,
while the drift voltage (HV1) was varied in steps of 25\,V in different ranges, depending on the anode voltage.
These experimental conditions, voltages used and the measured time resolution,
are summarized in Table~\ref{tab:LaserTimeResCOMPASS}.
In general, the lowest voltage used corresponds to a detector gain high enough to distinguish
the signal from the noise level (gain $\sim 10^5$), while the highest voltage is the maximum value for
which the detector operates in stable conditions (up to gains of $\sim 10^6$).
For each voltage configuration, more than $10^4$ events were recorded with the oscilloscope, and subsequently analyzed offline.

During the first data-taking campaign, the photocathode efficiency was less than 0.5\%,
which led us to derive the trigger directly from the PICOSEC signal,
in the interest of data collection efficiency.
Results are based on this first campaign as runs contain enough events for the analysis.
The PICOSEC trigger threshold varied between 10 and 90\,mV, as shown in Table~\ref{tab:LaserTimeResCOMPASS},
leading to a bias of the recorded PICOSEC pulse height spectrum.
In a later data-taking campaign, the photocathode was replaced to increase the signal efficiency up to $\sim$5\%;
this allowed deriving the trigger decision from the fast photodiode.
The detector was operated in the same voltage conditions as the data set with a trigger bias on the PICOSEC amplitude,
so they could be used to confirm that the charge distribution follows a Polya function~\cite{Zerguerras:2014yra},
as discussed in Sec.~\ref{sec:ResultsLaser}.

\subsection{Beam tests with 150\,GeV muons}
\label{sec:BeamSetup}
\begin{figure}[htb!]
\centering
\includegraphics[width=0.90\textwidth]{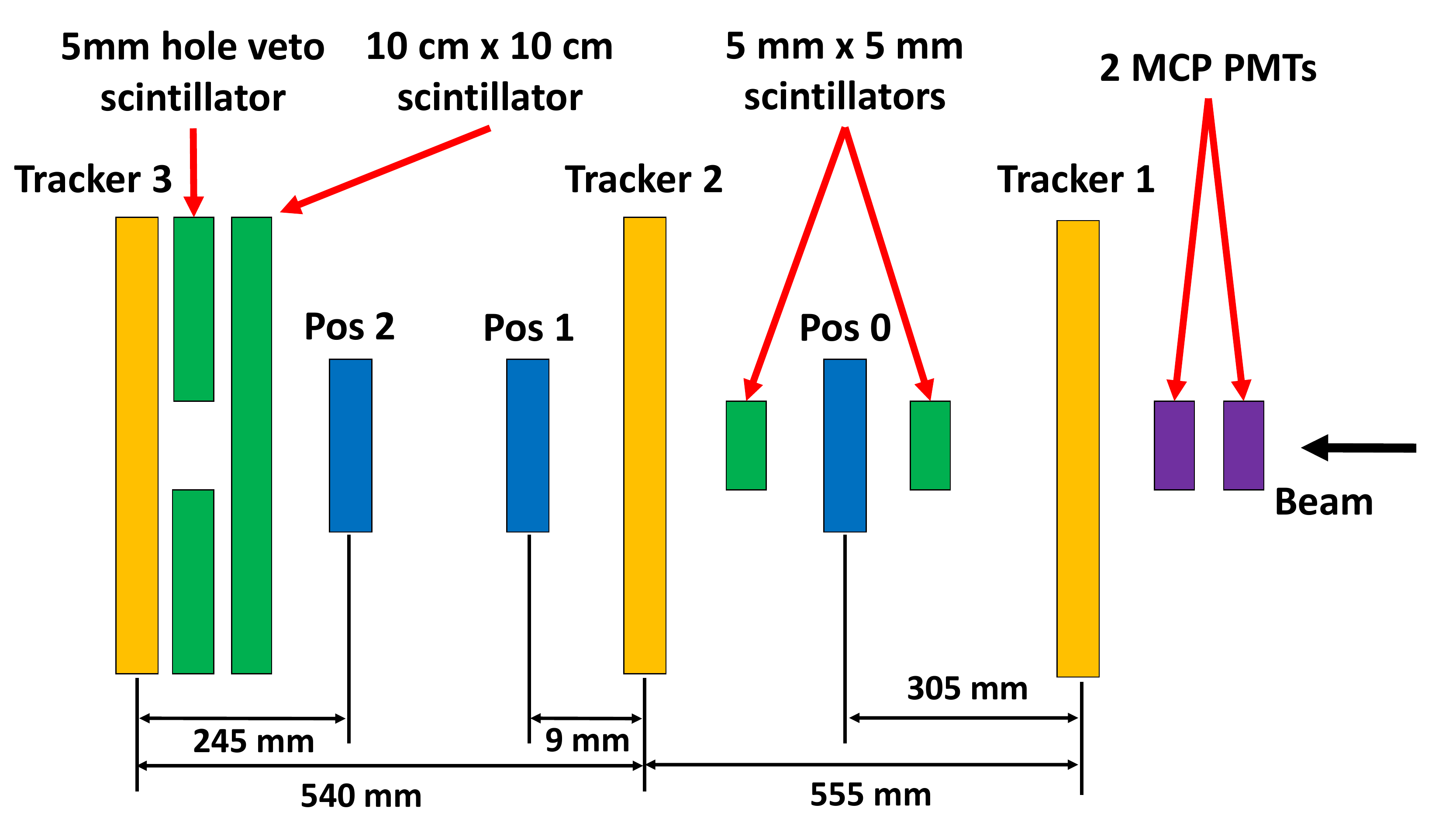}
\caption{Layout of the experimental setup (not to scale) during the beam tests. 
The incoming beam enters from the right side of the figure; events are triggered by the coincidence of 
two $5\times 5$\,mm$^2$ scintillators in anti-coincidence with a ``veto'' scintillator. 
Three GEM detectors provide tracking information of the incoming charged particles,
and the timing information is measured in three PICOSEC detectors (Pos0, Pos1, Pos2).
Details are given in the text.}
\label{fig:BeamSchema}
\end{figure}

The time response of the detector to 150\,GeV muons was measured
during several beam periods in 2016 and 2017 at the CERN SPS H4 secondary beamline.
The experimental setup (Fig.~\ref{fig:BeamSchema})
allows the characterization of up to three PICOSEC detectors, situated at the positions Pos0, Pos1 and Pos2.
Two trigger scintillators of $5 \times 5$\,mm$^2$ operate in anti-coincidence
with a veto scintillator whose aperture (hole) matches the same area.
This trigger configuration efficiently selects muons
that don't undergo scattering and suppresses triggers from particle showers.
One Hamamatsu MCP PMT\footnote{Model R3809U-50: \url{https://www.hamamatsu.com/jp/en/R3809U-50.html}} is used as time reference;
its entrance window (3\,mm thick quartz) is placed perpendicular to
the beam and serves as Cherenkov radiator.
From the time jitter between two identical MCPs we determine the reference time accuracy of 5.4$\,\pm$\,0.2\,ps.
A telescope of three tracking GEM detectors with two-dimensional strip readout
is used to reconstruct the trajectory of each muon
with a combinatorial Kalman filter based algorithm, and to determine its impact position at each detector.

The PICOSEC (and MCP reference) waveforms are recorded in the beam tests
in the same way as in the laser tests using an oscilloscope with 20\,GSamples/s and 2.5\,GHz bandwidth,
while the tracking (GEM detector) data are recorded simultaneously in an APV25 based SRS DAQ~\cite{Martoiu:2013sm}.
To ensure event alignment in the two DAQ systems,
the internal SRS event number is sent as a bit stream to one oscilloscope channel.
The DAQ trigger is generated by the scintillators.

Each PICOSEC detector is operated at different anode and drift voltages,
which are respectively scanned in steps of 25\,V.
As in the case of laser tests, the minimum and maximum drift voltages
respectively correspond to the cases where the signal is distinguishable from the noise level
and the detector operates in stable conditions.
However, the detector gain ($10^4-10^5$) was lower than in laser tests because
the initial number of electrons was higher.
For a fixed anode voltage, the drift voltage was $\sim$100\,V lower than in laser tests.
For each voltage configuration,
more than 4000 events are recorded with the oscilloscope, and subsequently analyzed offline.
During periods without beam or long accesses,
each detector is illuminated with a UV lamp to measure the response to single photoelectrons at different voltage settings.
This information is later analyzed to estimate the mean number of photoelectrons produced by muons during beam tests.

\subsection{Waveform analysis}
\label{sec:PulseAnalysis}
In this section, we briefly describe the analysis performed on both laser and beam test data.
For each PICOSEC signal, the baseline offset and noise level are determined using the 75\,ns precursor of the pulse.
Then, the ``electron-peak'' amplitude ($V_\mathrm{max}$) is defined
as the difference between the highest point of the waveform and the baseline.
For the timing measurement, a Constant Fraction (CF) method based on a sigmoid function is used
to minimize the contribution of the noise. Other algorithms to determine the CF have been used with similar results.
In this approach, a sigmoid function is fit to the leading edge of the electron-peak. This function is defined as
\begin{equation}
  V(t) = \frac{P_0}{1 + \exp\left(-P_2 \times (t - P_1)\right)} + P_3
  \label{eq:sigmoidfit}
\end{equation}
where $P_0$ and $P_3$ are respectively the maximum and the minimum values,
$P_1$ is the inflection time (i.e. where the slope changes derivative),
and $P_2$ quantifies the speed of the sigmoid change (i.e. is correlated to the signal risetime).
The time corresponding to a 20\% CF is calculated as follows:
\begin{equation}
  t_{z} = P_1 - \frac{1}{P_2} \ \log\Bigg[\frac{P_0}{0.2 \times V_{\mathrm{max}} - P_3} - 1\Bigg]
  \label{eq:sigmoidrep}
\end{equation}

For the photodetectors used as the time reference (i.e. MCP, or PD0 in the case of laser tests)
a simpler approach is applied as signals are almost immune to any source of noise:
after the calculation of the pulse baseline and amplitude,
a cubic interpolation between four points around CF=20\% is used to extract with better precision
the temporal position of the signal.
The ``Signal Arrival Time'' (SAT)
is then defined as the difference between the PICOSEC CF time and that of the reference detector,
as illustrated in Fig.~\ref{fig:PicosecSignal}.

The ``electron-peak charge'' is defined as
the integral of the waveform between the start and the end points,
defined as the first points situated before and after
the maximum whose amplitude is less than one standard deviation away from the baseline offset.
For those pulses with no clear separation between the electron-peak and the ion-tail,
the end point has been alternatively defined as the time when the pulse derivative changes sign.
The resulting value is then transformed to Coulombs, using the input impedance of\,$50$\,Ohm.
The measured electron-peak charge-to-amplitude ratio is 0.0033\,pC/mV.

\section{Laser test results}
\label{sec:ResultsLaser}
Two aspects of the PICOSEC time response in the laser measurements are discussed below.
Firstly, we discuss the  dependence of the time response on signal amplitude
as this dependence (particularly concerning the role of the drift field
and fluctuations in the preamplification at a given field) elucidates the physical origin of the PICOSEC time resolution.
Secondly, we convolute this amplitude dependence with the actual amplitude distribution corresponding to a single photoelectron.
Using this convolution, i.e. the full ``single photoelectron time response'',
we can then estimate the PICOSEC response for the case
of many photoelectrons produced in the Cherenkov signal from 150 GeV muons discussed in the next section.

\begin{figure}[htb!]
\centering
\includegraphics[width=0.49\textwidth]{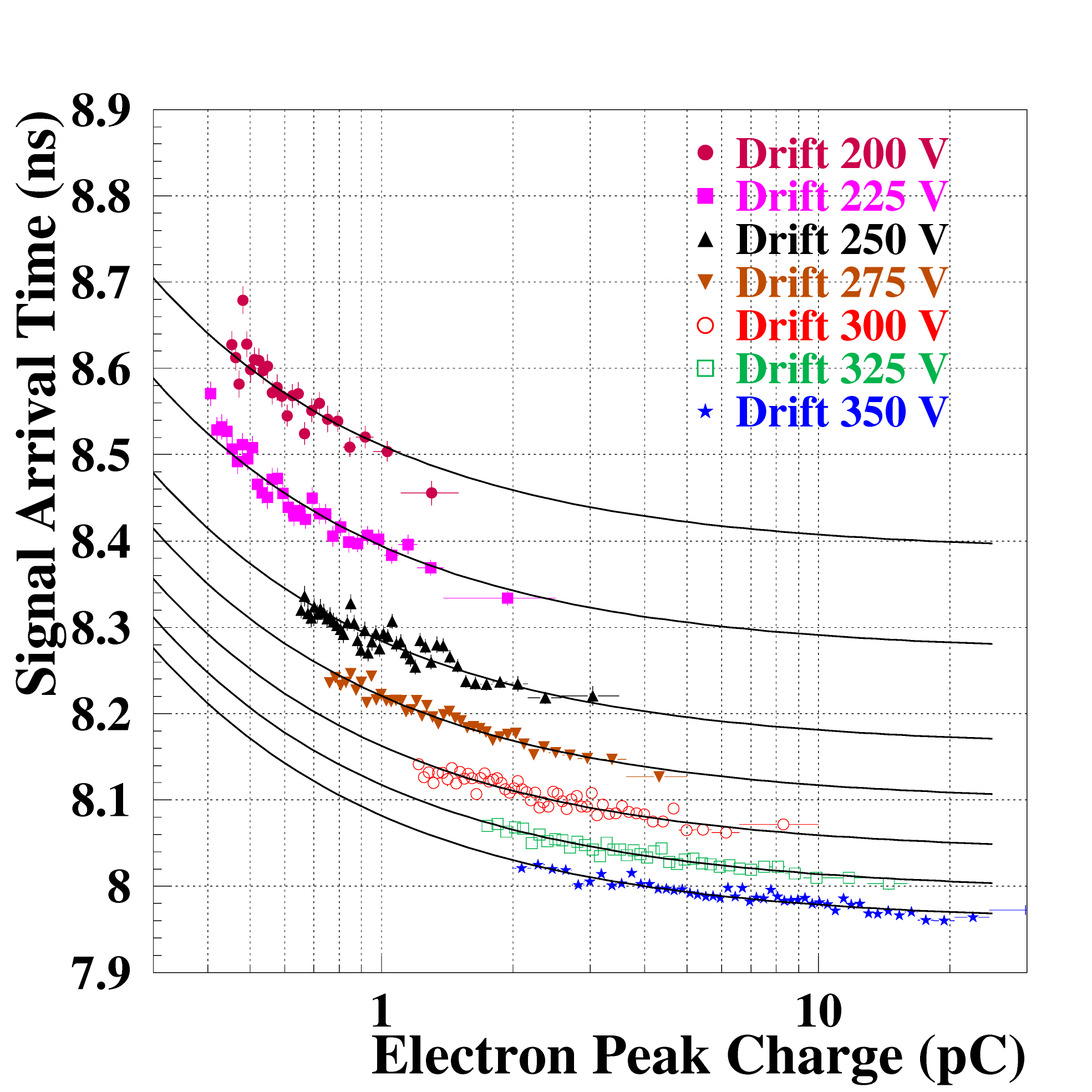}
\includegraphics[width=0.49\textwidth]{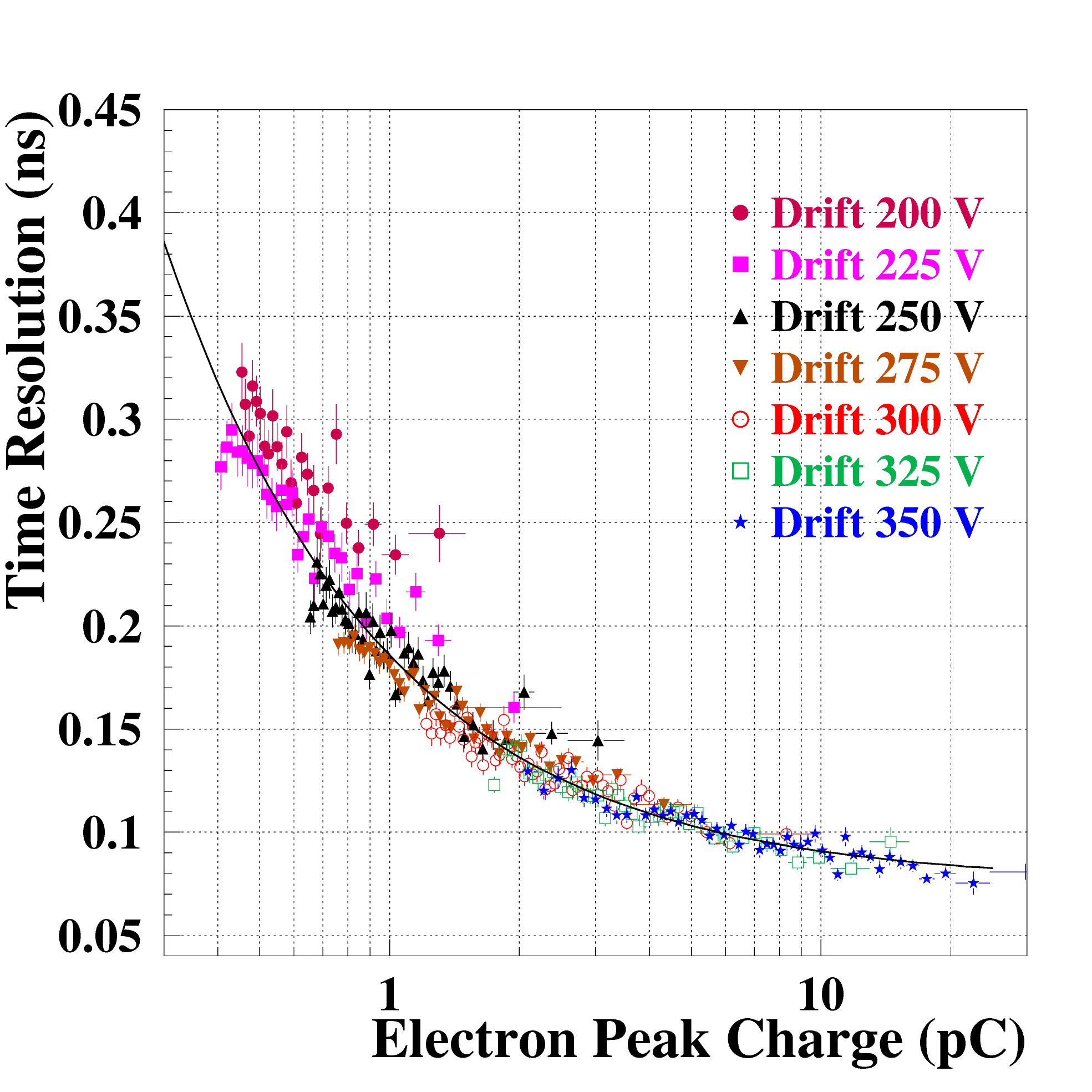}
\caption{Laser test:
Mean of the SAT values (left) and time resolution (right)
as a function of the electron-peak charge in case of single photoelectron data,
for an anode voltage of 525\,V and drift voltages between 200 and 350\,V.
The solid curves in the left distribution are the result of fitting
the functional form (see text, Eq.~\ref{eq:slewing}) to the experimental points
for each drift voltage, with the constraint that the parameters \textit{b} and \textit{w}
must be the same for all drift voltages.
Meanwhile, the solid curve in the right distribution is the result
of fitting the same equation to all experimental points, without any distinction of the drift voltage.
Statistical uncertainties are shown.}
\label{fig:DelayTResCharge}
\end{figure}

Since the experimental data on the SAT resolution approximately follow a Gaussian time distribution, we could simply report
the standard deviation as the time resolution of the PICOSEC detector.
However, there is a small tail at high SAT values, due to small charge (or amplitude) signals with late arrival time,
which accounts for a small fraction of the total events.
This results in a correlation between the SAT and the electron-peak charge (or amplitude).
This correlation is quantified for each voltage setting by the sample of PICOSEC signals in narrow ranges
of electron-peak charge and fitting with a Gaussian distribution the corresponding SAT values.
A typical dependence of the resulting mean and standard deviation values on the electron-peak charge
is shown in Fig.~\ref{fig:DelayTResCharge}. For both variables, there is a decrease with the charge,
which can be described by the following parametric function:

\begin{equation}
y = \frac{b}{x^w} + a
\label{eq:slewing}
\end{equation}

where $y$ is either the mean or standard deviation of the SAT,
$x$ is the electron-peak charge; and \textit{a}, \textit{b} and \textit{w} are three free parameters.
For the mean values, this function is used to fit the experimental points
with the constraint that the parameters \textit{b} and \textit{w}
must be the same for all datasets
with the same anode voltage, while the parameter $a$ could take different values for each drift voltage setting.
As shown in Fig.~\ref{fig:DelayTResCharge} (left), this multiple fit function works well.
The values for the standard deviation of the SAT (i.e. the time resolution) follow a common curve
for all data with the same anode voltage and can be thus fitted with the same function of Eq.~\ref{eq:slewing}.
Fig.~\ref{fig:DelayTResCharge} (right) demonstrates that the fit works well.
The data points deviate slightly from the curve at
low electron-peak signal amplitudes. Overall, Fig.~\ref{fig:DelayTResCharge} (right) shows an
improved time resolution for an earlier onset of the avalanche (i.e. a
higher signal charge). In summary, these results indicate: a)
a decrease of the parameter $a$ with drift voltage (c.f. Fig.~\ref{fig:DelayTResCharge} (left)),
that reflects the dependence of the electron drift velocity on
drift field, and that b) the time resolution properties of the PICOSEC
detector are described by a single function and are mainly determined by
the electron-peak charge.

In fact there are two physical parameters of the drift region which could, through their dependence on
the applied drift voltage, affect the timing performance: a) the
longitudinal diffusion coefficient (i.e.  $\sigma_\mathrm{t} \sim D^{1/2}_\mathrm{L} / v_\mathrm{drift}$), and b)
the mean free path to the first ionizing collision. The observed scaling
with the pulse amplitude for fixed anode voltage suggests that the
improvement of the time resolution is driven by the latter effect (i.e.
“b)” ), while the contribution from the variation in longitudinal diffusion is
less significant in this regime.

\begin{figure}[htb!]
\centering
\includegraphics[width=0.70\textwidth]{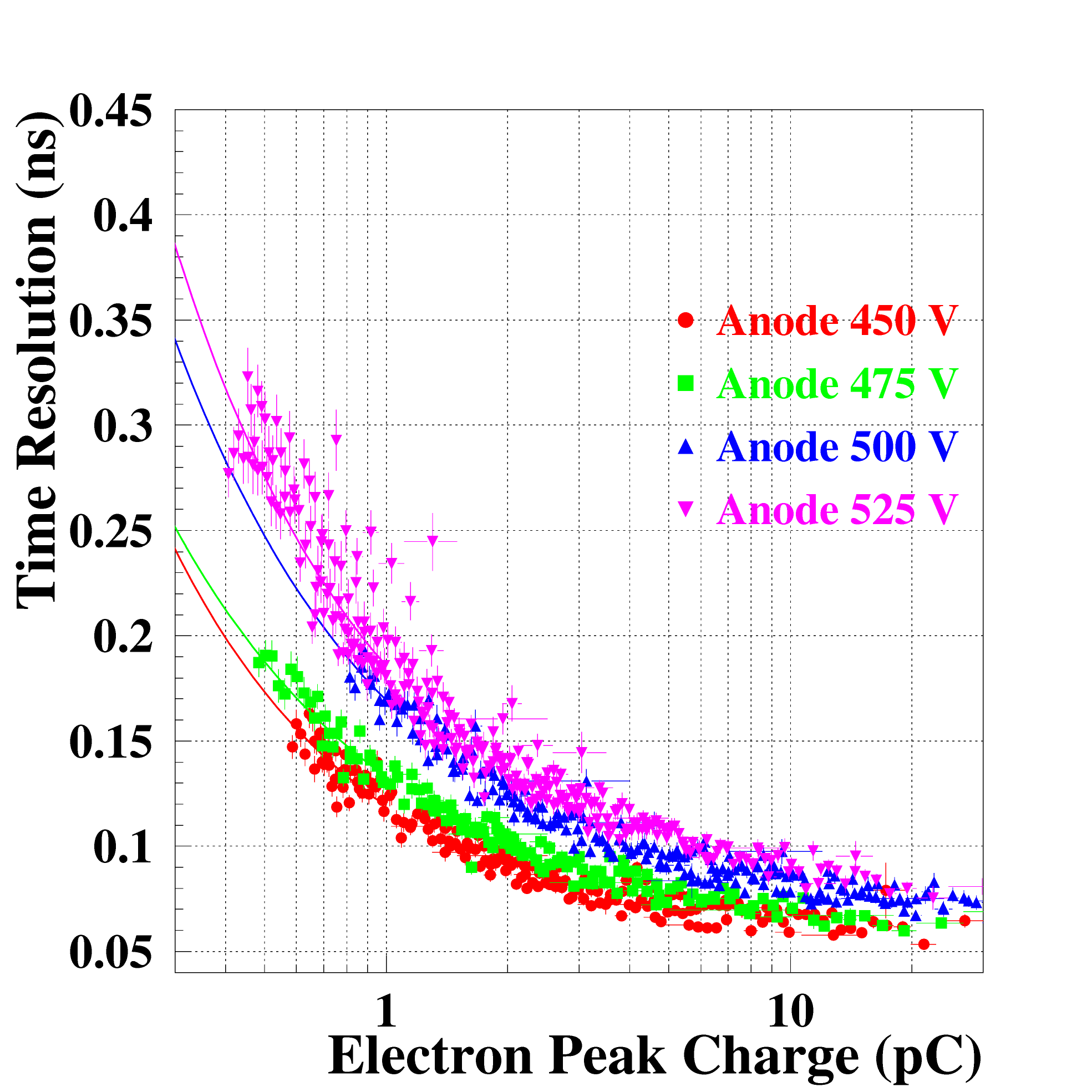}
\caption{Laser test: Dependence of the time resolution on the electron-peak charge
for anode voltages of 450\,V (red circles, drift voltages between 300 and 425\,V),
475\,V (green squares, drift voltages between 300 and 400\,V),
500\,V (blue triangles, drift voltages between 275 and 400\,V),
and 525\,V (magenta inverted triangles, drift voltages between 200 and 350\,V).
The continuous lines are the result of fitting the functional form of Eq.~\ref{eq:slewing}
to the experimental points for the same anode voltage  (see text).
Statistical uncertainties are shown.}
\label{fig:TResVsAnode}
\end{figure}

Yet another indication of the dominant role of the drift region can be derived from Fig.~\ref{fig:TResVsAnode},
where the dependence of the time resolution on electron-peak charge is shown for different anode voltages.
To preserve the same level of signal amplitudes at lower anode voltages,
the drift fields have been correspondingly increased.
Signals with the same electron-peak charge and lower anode voltage
(and thus necessitating a higher preamplification)
show a better time resolution, i.e.,
pulses with a higher preamplification gain have better timing properties than those with a higher amplification gain.

\begin{figure}[htb!]
\centering
\includegraphics[width=0.90\textwidth]{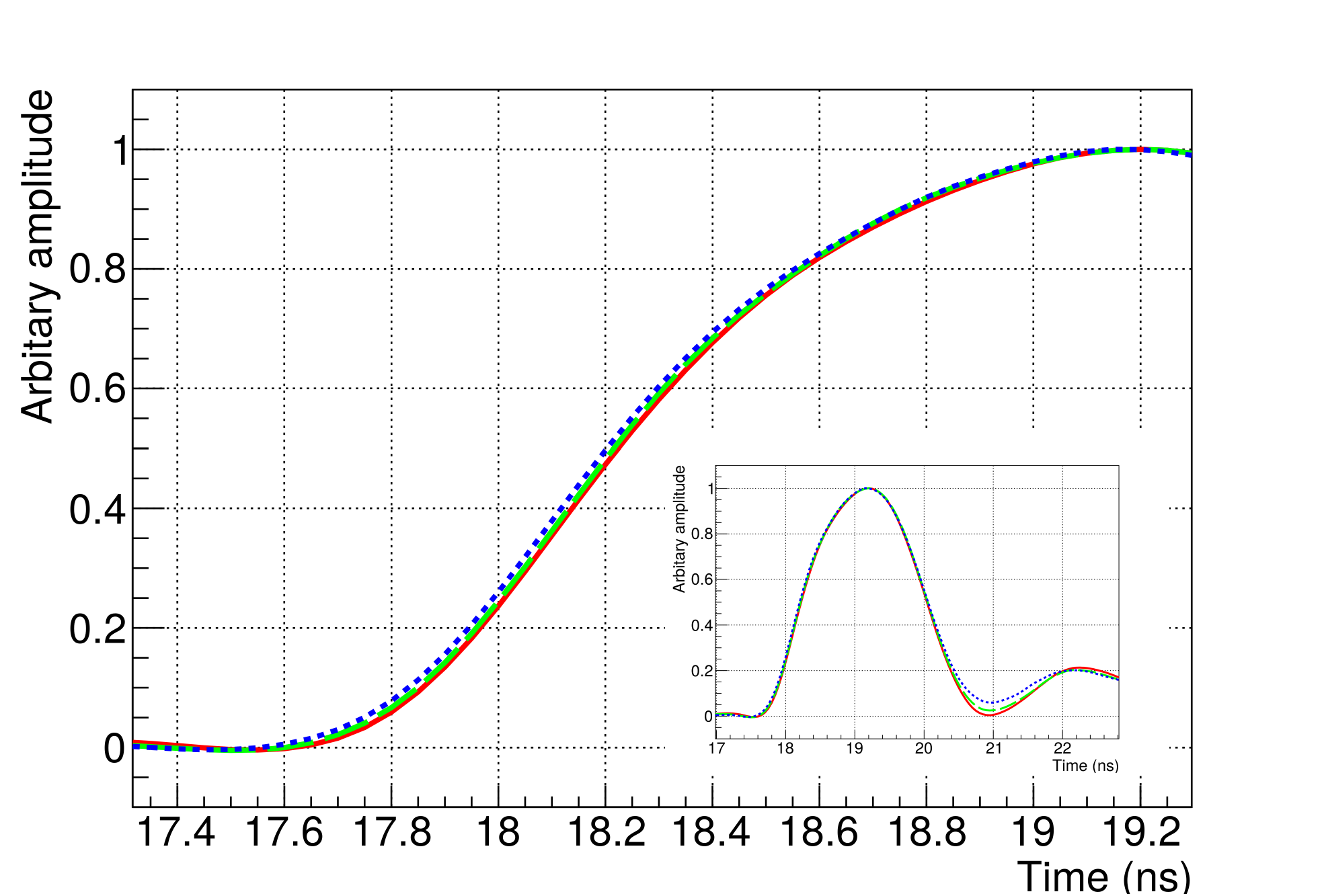}
\caption{Laser test: Average of the electron-peak shape normalized to unity for electron-peak charges
of 1.0-1.1\,pC (continuous red line), 2.0-2.5\,pC (segmented green line) and 3-4\,pC (dashed blue line).
The figure shows a zoom to the leading edge, while the inset shows the complete electron-peak component.
The detector was operated at an anode voltage of 450\,V and a drift voltage of 350\,V
with the ``COMPASS gas'' at 1\,bar absolute pressure.}
\label{fig:AveragePulse}
\end{figure}

The CF algorithm discussed above is used to eliminate the expected correlation between signal amplitude and SAT 
observed for signals with similar shapes but different amplitudes (known as ``time walk correction''~\cite{Delagnes:2016hdo})
normally observed when timing is derived from a fixed threshold.
Nevertheless, there are also well known
examples where both amplitude and signal risetime can vary from pulse to pulse,
requiring  ``amplitude and risetime correction'' for SAT determination.
We also considered this hypothesis since the time resolution varies by several hundreds of picoseconds,
for different signal amplitudes \textemdash even with the CF method.
However, as shown in Fig.~\ref{fig:AveragePulse}, the average electron-peak shape remains essentially identical
for different electron-peak charges.
For this reason, the correlation between electron-peak charge and the signal arrival time
observed in Fig.~\ref{fig:AveragePulse} must be a consequence of the physical mechanism generating the PICOSEC signal
rather than an artifact of the timing algorithm.

\subsection{Derivation of the overall ``single photoelectron time distribution function''}
\label{sec:singlephotoelectron}
\begin{figure}[htb!]
\centering
\includegraphics[width=0.48\textwidth]{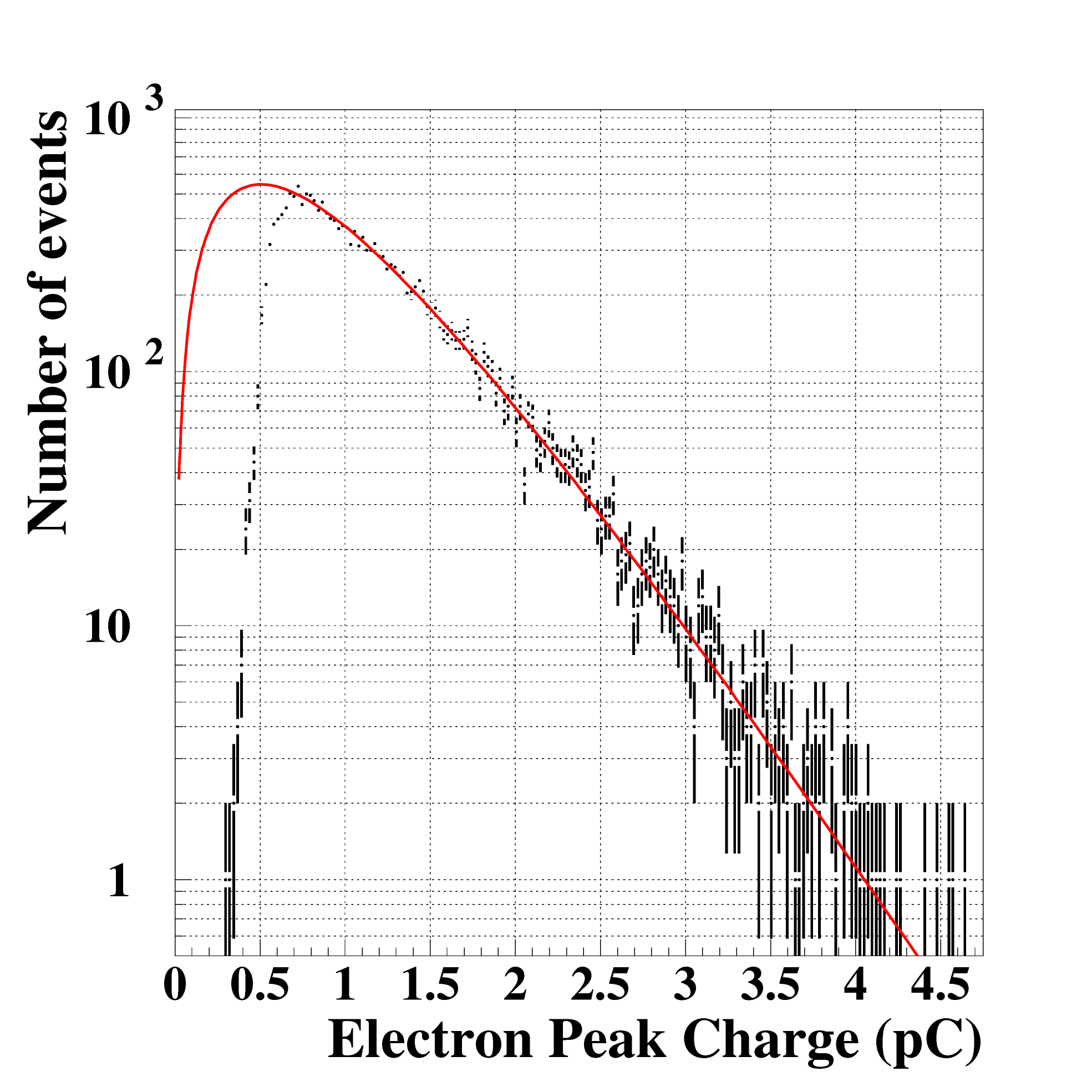}
\includegraphics[width=0.48\textwidth]{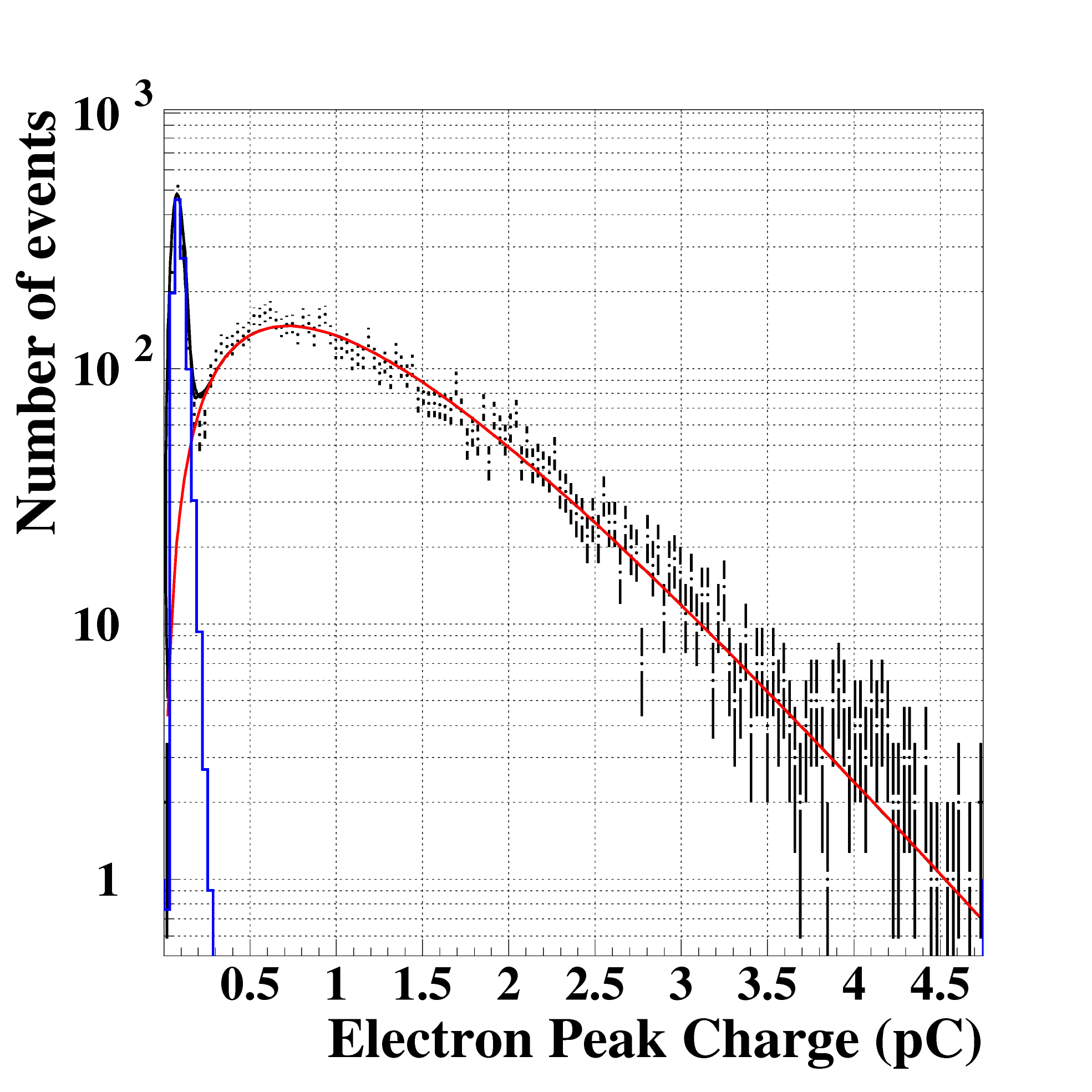}
\caption{Laser test: Two examples of the electron-peak charge distributions generated by single photoelectrons:
one biased by the PICOSEC detector threshold (left),
and another unbiased- using only the reference photodetector in the trigger chain (right).
The voltage settings in both cases are 450\,V for the anode and 350\,V for the drift.
In both cases, the charge distribution is fit by a Polya function (red line), and
with a separate noise contribution (blue line in the right plot).
Statistical and systematic uncertainties are shown.}
\label{fig:PolyaFitting}
\end{figure}

As described in Sec.~\ref{sec:LaserSetup}, data are collected with an electronic trigger
generated by the PICOSEC detector for part of the dataset. The threshold level was in some cases
high in comparison to the Root Mean Square (RMS) baseline noise (typically $\sim$2.5\,mV),
as detailed in Table~\ref{tab:LaserTimeResCOMPASS}. Supposing that the derived dependence of the mean
and standard deviation of the SAT with the electron-peak charge are also valid for pulse amplitudes lower
than the threshold, the time resolution of the PICOSEC detector signal at a given operating point
is estimated by the equation:

\begin{equation}
\sigma^2 = \sum^n_{i = 1} a^2_i \ \sigma^2_i 
+ \sum^n_{i = 1} \sum^n_{j = i + 1} a_i \times a_j \times \bigg(\sigma^2_i + \sigma^2_j + \left(\mu_i - \mu_j\right)^2\bigg)
\label{eq:slewthrcorrection}
\end{equation}

where $\mu_i$ and $\sigma_i$ are the mean and standard deviation of the SAT
in an interval $i$ ($i=1,n$) of the electron-peak charge ($Q_i$), 
and $a_i$ is the probability density function (PDF) for a given charge $Q_i$,
i.e. $a_i = A(Q_i)$, and $\sum^n_{i = 1} a_i = 1$.
The term $(\mu_i - \mu_j)$ removes the difference in SAT at different electron-peak charges,
caused by the measured correlation between these two variables.
For all cases, the minimum electron-peak charge $Q_1$ is set to 0.033\,pC (i.e. 10\,mV in amplitude),
equivalent to four times the typical RMS baseline noise.
Meanwhile, the $A(Q)$ PDFs are obtained by fitting each electron-peak charge distribution
by a Polya function~\cite{Zerguerras:2014yra} which is expressed as:

\begin{equation}
A(Q|N,Q_e,\theta) = \frac{(\theta + 1)^{N (\theta + 1)}}{\Gamma(N (\theta + 1))}
\left(\frac{Q}{Q_e}\right)^{N (\theta + 1) -1} \exp\Bigg[- (\theta + 1) \frac{Q}{Q_e}\Bigg]
\label{eq:Polyafunction} 
\end{equation}

where $Q_e$ is the mean charge per single photoelectron, $N$ is the number of photoelectrons,
and $\theta$ is the Polya shape parameter.
This function describes well the single electron-peak charge response of the PICOSEC detector ($N = 1$),
as shown in Fig.~\ref{fig:PolyaFitting},
including also the dataset without a PICOSEC trigger threshold bias (Fig.~\ref{fig:PolyaFitting}, right).
In each fit, bin sizes and fitting regions were varied in order to estimate the systematic errors,
which were then combined with the statistical uncertainties.

\begin{figure}[htb!]
\centering
\includegraphics[width=0.70\textwidth]{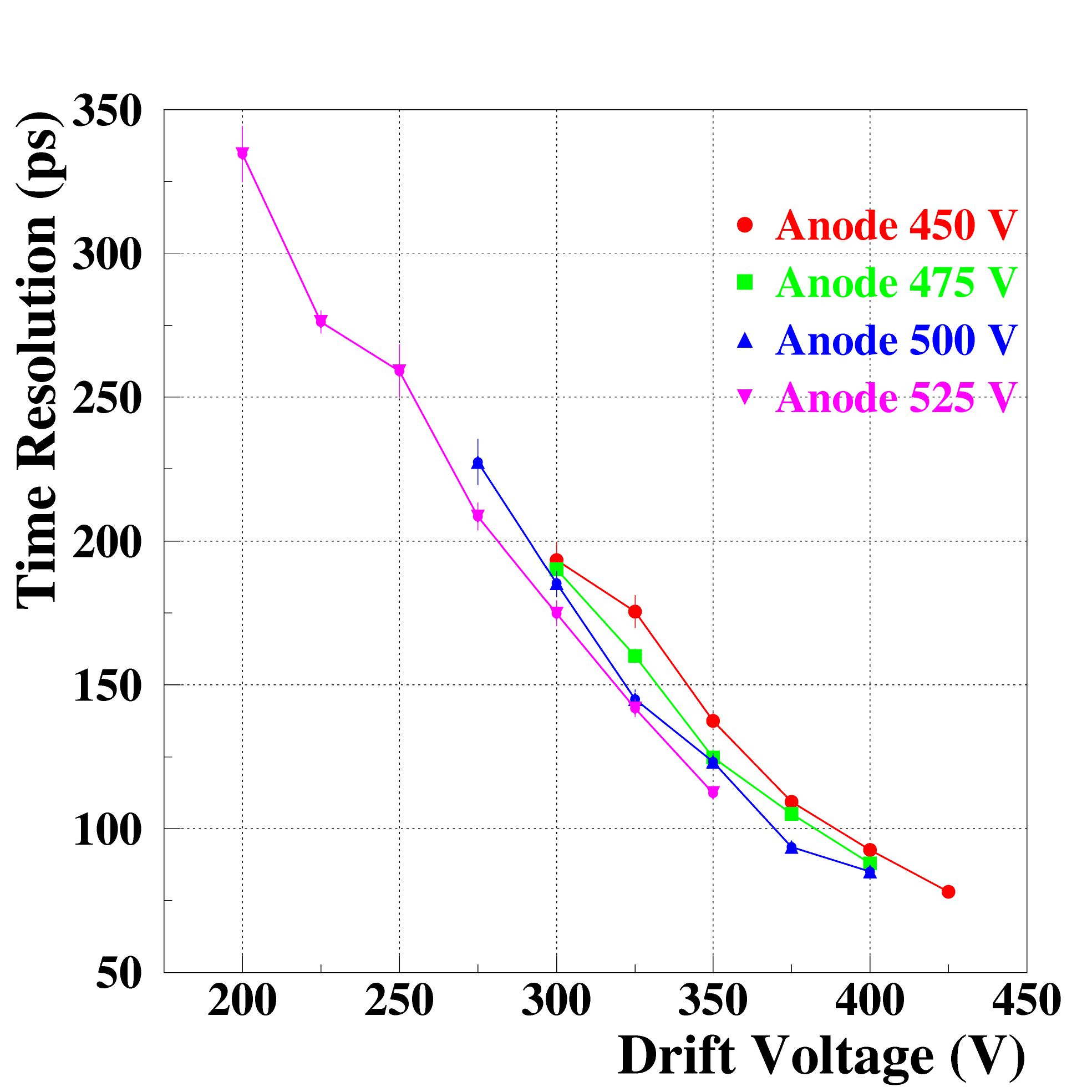}
\caption{Laser test: Dependence of the corrected time resolution
on the drift voltage for anode voltages of 450\,V (red circles), 475\,V (green squares),
500\,V (blue triangles) and 525\,V (magenta inverted triangles).
Statistical uncertainties are shown.}
\label{fig:TResVsDrift}
\end{figure}

The raw and corrected time resolution values of the PICOSEC detector
for single photoelectron detection at the different operation settings
are shown in Table~\ref{tab:LaserTimeResCOMPASS}.
Two uncertainties are included in the calculation:
the uncertainty in the parametrization of the mean and of the standard deviation of the SAT
with the charge, and the uncertainty in the Polya parametrization.
The best measured value of the time resolution is 76.0 $\pm$ 0.4\,ps,
which is obtained for the lowest applied anode voltage (450\,V).
Meanwhile, as can be seen from the dependence of the time resolution on
the drift and anode voltages (Fig.~\ref{fig:TResVsDrift}),
the time resolution is better for higher drift voltage.
We did not explore the whole parameter space but a further improvement is expected
if the drift voltage can be further increased,
while keeping the gain almost constant (by reducing the anode voltage correspondingly).
In fact, a simulation of the detector response~\cite{Paraschou:2017kp} has shown that
the detector time jitter is mainly defined by the drift (pre-amplification) stage,
while the contribution of the amplification stage is negligible.

\begin{table*}[htb!]
\centering
\caption{Laser tests:
The experimental conditions, voltages used and the time resolution (raw and corrected values) of the PICOSEC detector,
operated with the ``COMPASS gas'' at 1\,bar absolute pressure. The threshold is the amplitude of the smallest signal recorded
at the trigger level
and the RMS is the standard deviation of the baseline.
The raw time resolution values are estimated using the CF algorithm,
while the corrected values are estimated after correcting 
the correlation of the SAT with the electron-peak charge, as discussed in detail in Sec.~\ref{sec:singlephotoelectron}.} 
\begin{tabular}{cc|cc|cc}
\hline
Anode & Drift & Threshold & RMS  & \multicolumn{2}{c}{Time resolution (ps)}\\
(V)   & (V)   & (mV)      & (mV) & Raw & Corrected\\
\hline
450 & 300 & 14.3 $\pm$ 0.3 & 2.5 $\pm$ 0.2 & 164.0 $\pm$ 4.2 & 184.6 $\pm$ 1.4\\
    & 325 & 15.8 $\pm$ 0.3 & 2.5 $\pm$ 0.2 & 147.0 $\pm$ 4.0 & 169.5 $\pm$ 1.1\\
    & 350 & 16.1 $\pm$ 0.4 & 2.5 $\pm$ 0.2 & 121.6 $\pm$ 1.8 & 140.1 $\pm$ 1.0\\
    & 375 & 26.9 $\pm$ 0.6 & 2.5 $\pm$ 0.2 &  88.4 $\pm$ 0.5 & 108.7 $\pm$ 0.6\\
    & 400 & 37.5 $\pm$ 1.1 & 2.9 $\pm$ 0.2 &  77.0 $\pm$ 0.5 &  90.3 $\pm$ 0.5\\
    & 425 & 79.4 $\pm$ 2.7 & 5.6 $\pm$ 0.3 &  69.5 $\pm$ 0.6 &  76.0 $\pm$ 0.4\\
\hline
475 & 300 & 11.5 $\pm$ 0.3 & 2.5 $\pm$ 0.2 & 180.0 $\pm$ 6.0 & 187.8 $\pm$ 0.8\\
    & 325 & 16.1 $\pm$ 0.4 & 2.5 $\pm$ 0.2 & 140.0 $\pm$ 1.0 & 160.3 $\pm$ 0.7\\
    & 350 & 30.3 $\pm$ 0.7 & 2.6 $\pm$ 0.2 &  90.7 $\pm$ 0.6 & 123.8 $\pm$ 1.0\\
    & 375 & 31.6 $\pm$ 1.1 & 2.6 $\pm$ 0.2 &  89.0 $\pm$ 0.6 & 105.3 $\pm$ 0.5\\
    & 400 & 44.6 $\pm$ 2.2 & 2.9 $\pm$ 0.2 &  79.1 $\pm$ 0.5 &  86.0 $\pm$ 0.3\\
\hline
500 & 275 & 20.6 $\pm$ 0.4 & 3.1 $\pm$ 0.3 & 175.0 $\pm$ 3.1 & 230.0 $\pm$ 3.0\\
    & 300 & 21.1 $\pm$ 0.5 & 3.4 $\pm$ 0.4 & 150.8 $\pm$ 1.8 & 186.0 $\pm$ 2.0\\
    & 325 & 30.6 $\pm$ 0.8 & 3.1 $\pm$ 0.2 & 115.8 $\pm$ 1.2 & 145.5 $\pm$ 1.0\\
    & 350 & 41.8 $\pm$ 1.2 & 3.4 $\pm$ 0.3 &  98.3 $\pm$ 0.9 & 121.2 $\pm$ 1.0\\
    & 375 & 87.9 $\pm$ 2.6 & 5.9 $\pm$ 0.3 &  85.3 $\pm$ 0.5 &  92.6 $\pm$ 0.6\\
    & 400 & 93.7 $\pm$ 4.7 & 5.7 $\pm$ 0.2 &  78.8 $\pm$ 0.5 &  83.8 $\pm$ 0.3\\
\hline
525 & 200 & 11.1 $\pm$ 0.2 & 2.6 $\pm$ 0.2 & 290.0 $\pm$ 7.0 & 337.5 $\pm$ 2.0\\
    & 225 & 11.1 $\pm$ 0.2 & 2.7 $\pm$ 0.2 & 261.8 $\pm$ 3.0 & 278.0 $\pm$ 1.2\\
    & 250 & 15.6 $\pm$ 0.3 & 2.6 $\pm$ 0.2 & 210.3 $\pm$ 3.0 & 254.2 $\pm$ 2.0\\
    & 275 & 15.7 $\pm$ 0.4 & 2.6 $\pm$ 0.2 & 180.2 $\pm$ 2.0 & 208.4 $\pm$ 1.0\\
    & 300 & 29.9 $\pm$ 0.6 & 2.7 $\pm$ 0.2 & 133.0 $\pm$ 1.4 & 174.8 $\pm$ 1.0\\
    & 325 & 41.9 $\pm$ 0.4 & 3.0 $\pm$ 0.2 & 111.9 $\pm$ 0.8 & 141.6 $\pm$ 0.7\\
    & 350 & 43.1 $\pm$ 1.6 & 3.0 $\pm$ 0.2 & 100.7 $\pm$ 0.9 & 110.5 $\pm$ 0.5\\
\hline
\end{tabular}
\label{tab:LaserTimeResCOMPASS}
\end{table*}

\section{Beam tests results with 150\,GeV muons}
\label{sec:ResultsBeam}
\begin{figure}[htb!]
\centering
\includegraphics[width=0.49\textwidth]{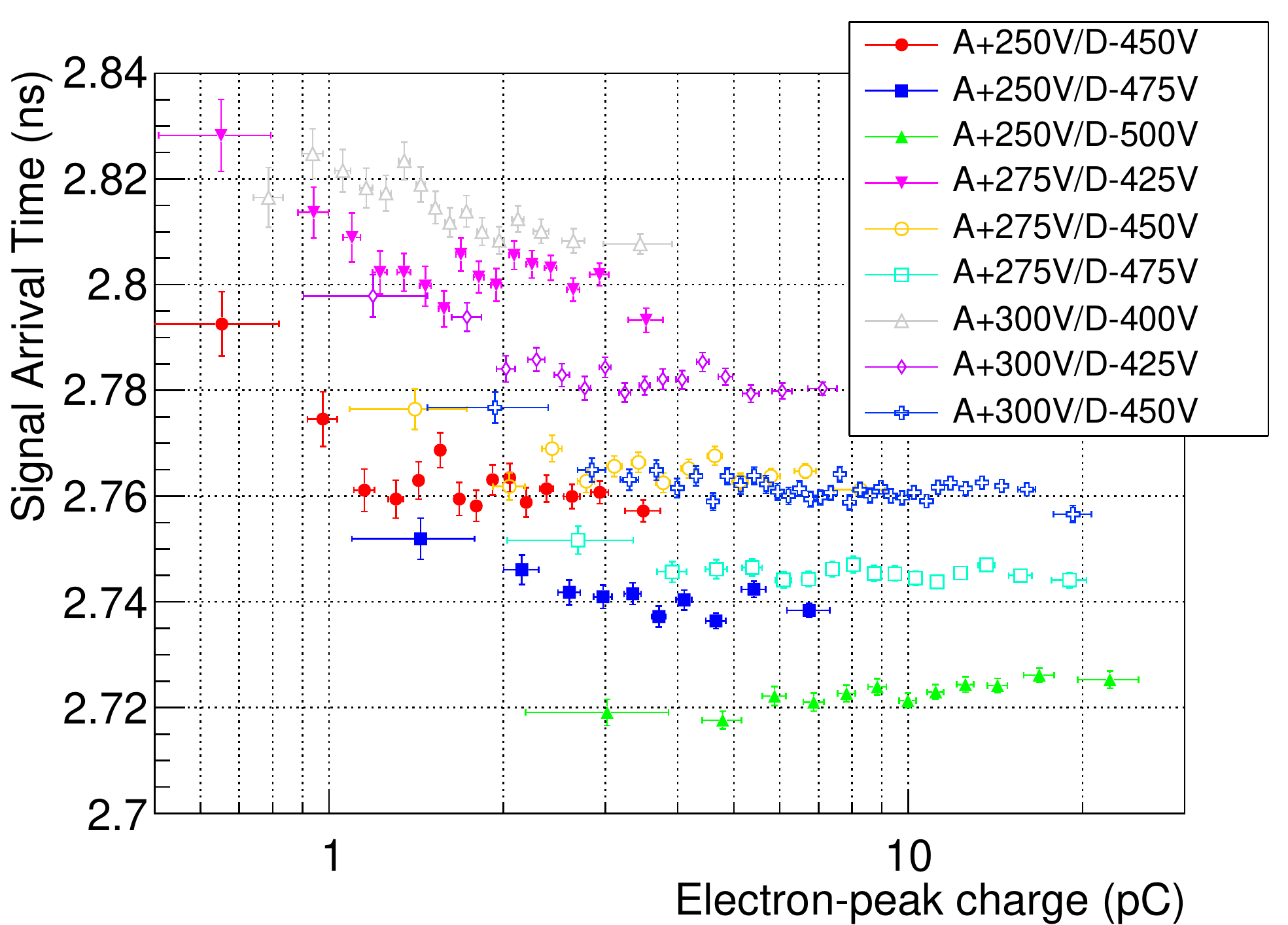}
\includegraphics[width=0.49\textwidth]{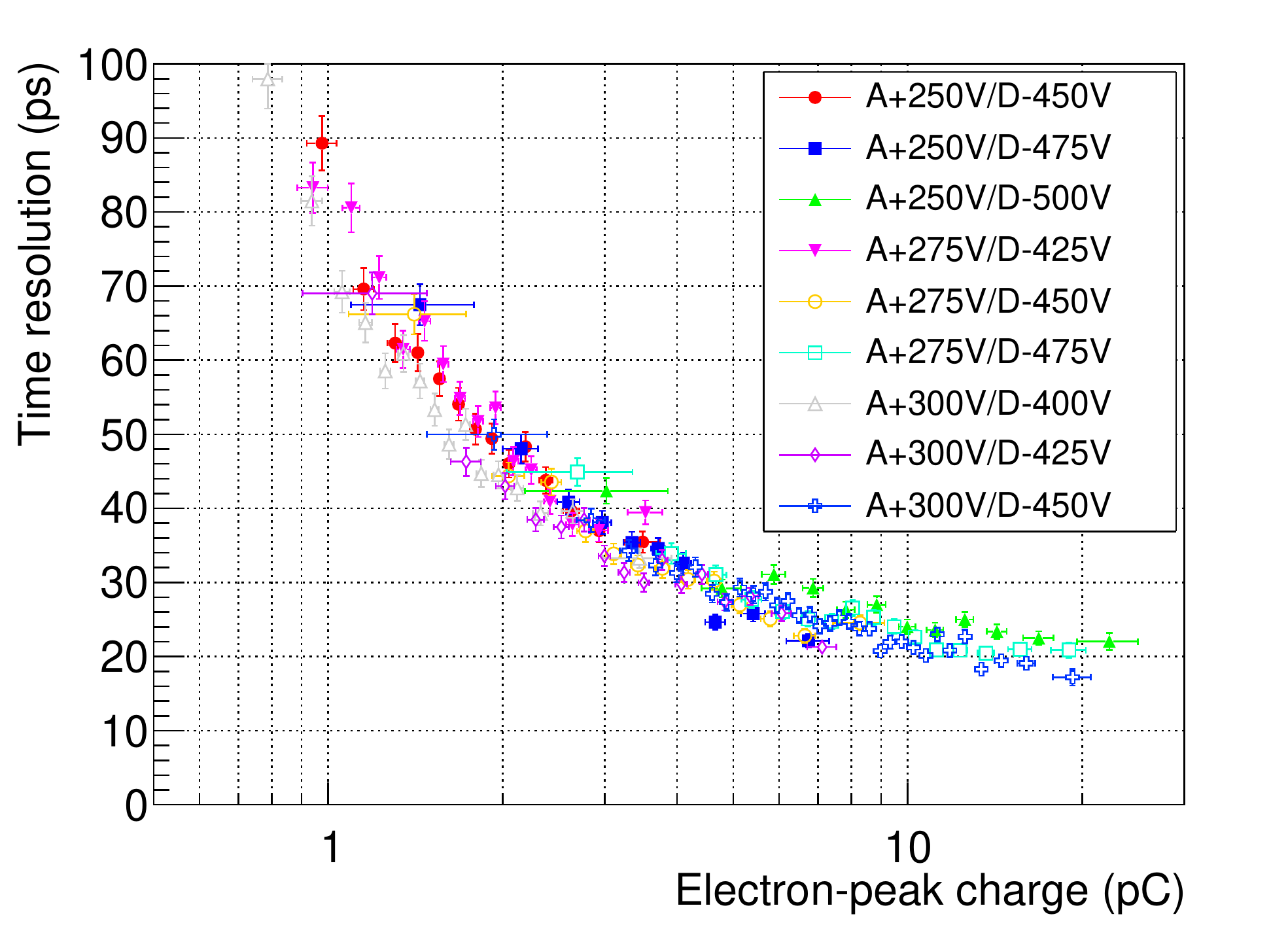}
\caption{Beam test: Dependence of the signal arrival time (left) and the time resolution (right)
on the electron-peak charge for 150\,GeV muons, for anode (A) voltages between 250\,V and 300\,V
and drift voltages (D) between 400\,V and 500\,V.
Statistical uncertainties are shown.}
\label{fig:DelayTResChargeBeam150GeV}
\end{figure}

The same analysis as in Sec.~\ref{sec:ResultsLaser} is applied to the SAT distributions
of 150\,GeV muons, as a correlation with the electron-peak charge is expected.
However, as shown in Fig.~\ref{fig:DelayTResChargeBeam150GeV} (left),
the mean of the SAT distribution is almost constant for each setting; this is explainable
by the high drift fields (and preamplification gains) at which the PICOSEC detector is operated.
Meanwhile, the time resolution decreases as the electron-peak charge increases (Fig.~\ref{fig:DelayTResChargeBeam150GeV}, right).

\begin{figure}[htb!]
\centering
\includegraphics[width=0.80\textwidth]{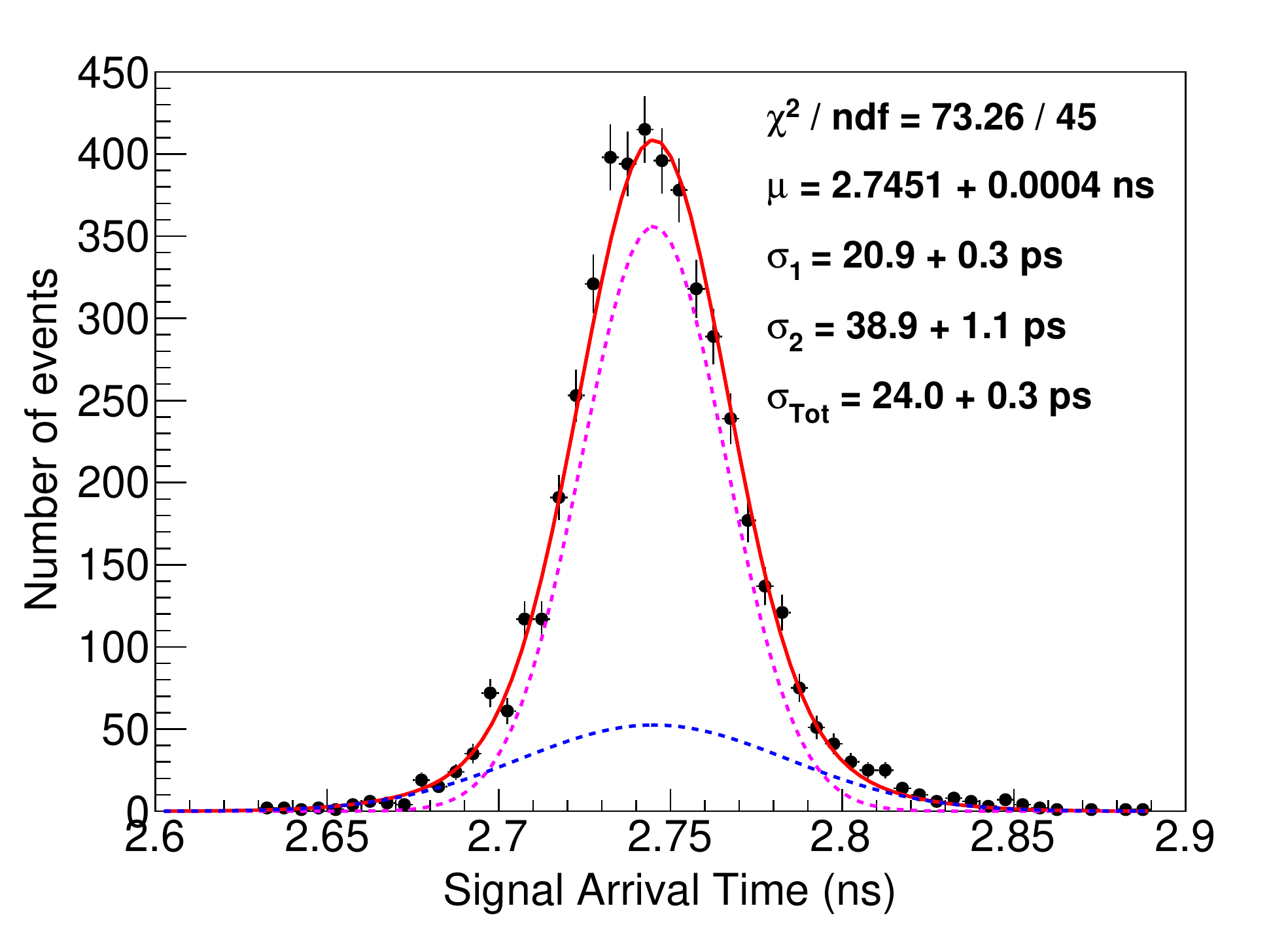}
\caption{Beam test: An example of the signal arrival time distribution for 150\,GeV muons, 
and the superimposed fit with a two Gaussian function (red line for the combination
and dashed blue and magenta lines for each Gaussian function),
for an anode and drift voltage of 275\,V and 475\,V, respectively.
Statistical uncertainties are shown.}
\label{fig:BeamSATBest}
\end{figure}

As the mean of the SAT distribution is almost independent of the electron-peak charge,
each SAT distribution generated by 150\,GeV muons is fit
by a two Gaussian distribution (both Gaussians centered at the same value)
and the time resolution is reported as the standard deviation
of the full distribution\footnote{A single Gaussian fit has also been used with similar results.}.
The time resolution results obtained are as low as 24.0 $\pm$ 0.3\,ps,
as shown in Fig.~\ref{fig:BeamSATBest} for anode and drift voltages of 275\,V and 475\,V, respectively.

\begin{figure}[htb!]
\centering
\includegraphics[width=0.80\textwidth]{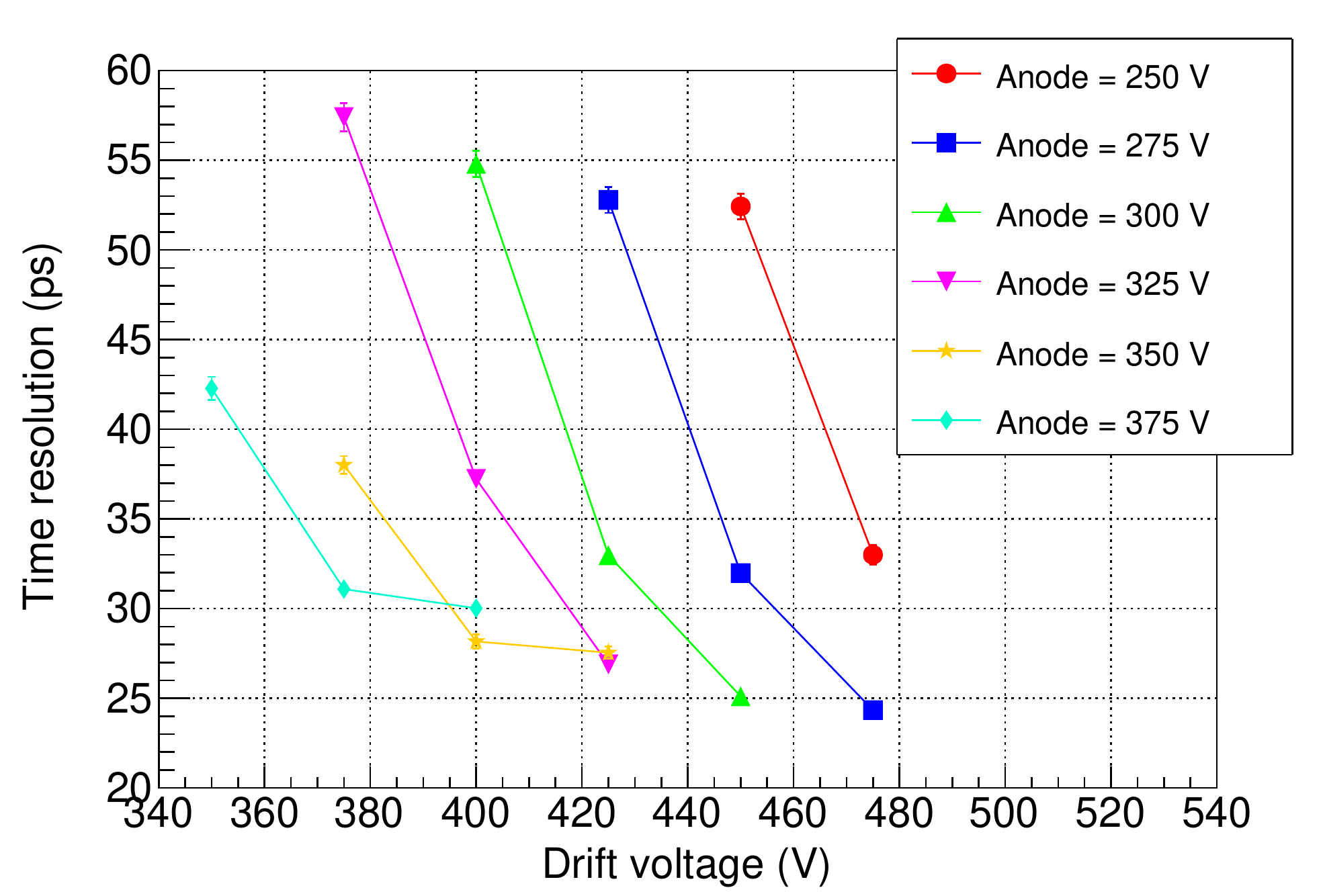}
\caption{Beam test: Dependence of the time resolution on the drift and anode voltage
for a PICOSEC detector irradiated by 150\,GeV muons. For each curve at a given anode voltage,
the maximum drift voltage corresponds to the maximum gain at which the detector can work in stable conditions.
Statistical uncertainties are shown.}
\label{fig:BeamTResVsDrift}
\end{figure}

From a scan over a wide range of voltage settings
we obtain the dependence of the time resolution
on the drift and anode voltages, as shown in Fig.~\ref{fig:BeamTResVsDrift}.
This figure clearly shows that the time resolution improves for higher drift voltages, 
while the gain is kept constant by reducing in the same proportion the anode voltage.
The optimal time resolution is reached for drift voltages of 450-475\,V,
which are the maximum settings at which the detector can be stably operated,
i.e. there is no discharge during the beam run.

\begin{figure}[htb!]
\centering
\includegraphics[width=0.80\textwidth]{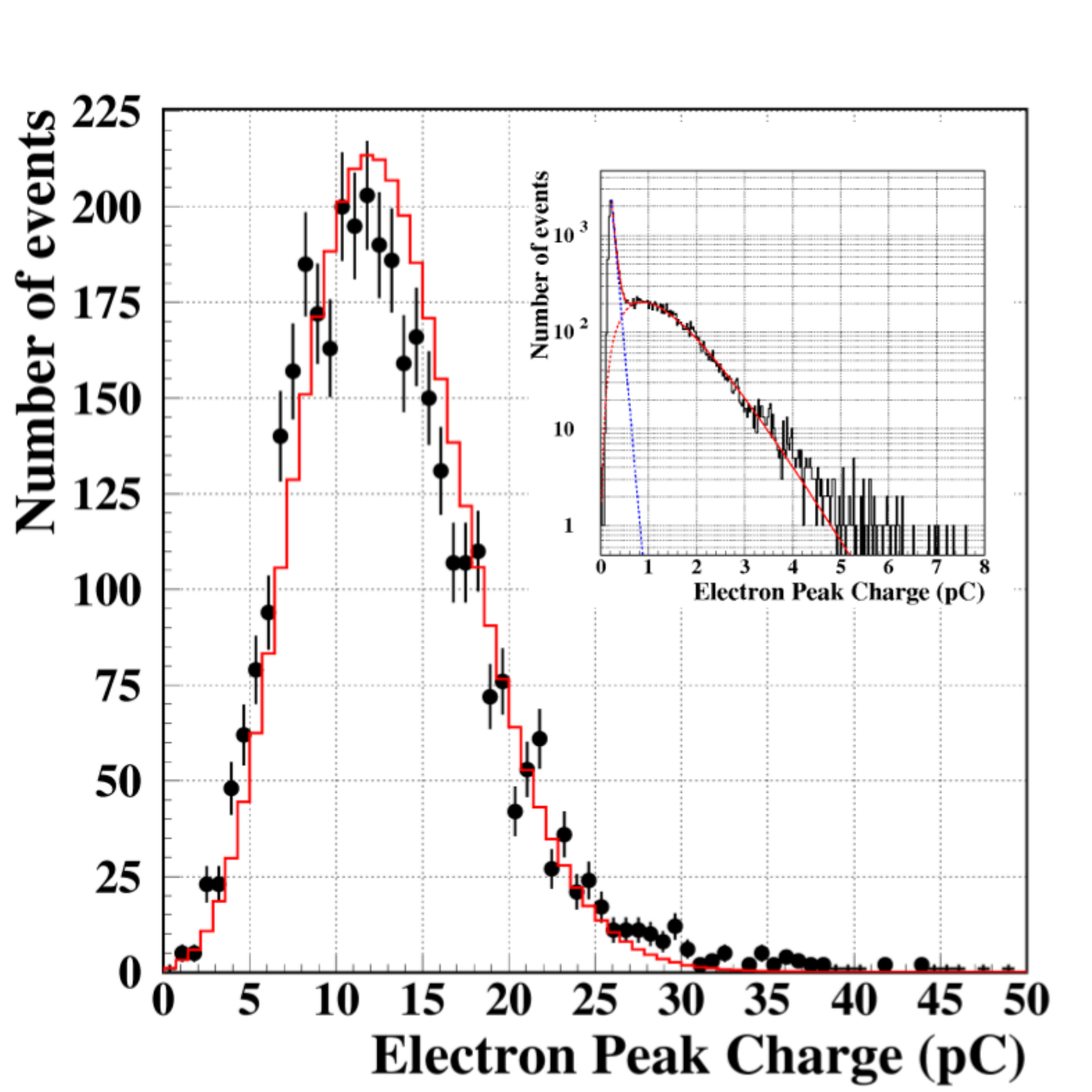}
\caption{Beam test: An example of the electron-peak charge distribution (black points) generated by 150\,GeV muons
and compared to the statistical prediction (red line) obtained from a maximum likelihood method.
Inset: the electron-peak distribution generated by a signal from the UV lamp (black points) is
fit by the single electron-peak distribution (red line) described by Eq.~\ref{eq:Polyafunction}
plus a noise contribution (blue line).
The settings are 275\,V for the anode and 475\,V for the drift voltages, in both cases.
Statistical uncertainties are shown.}
\label{fig:PolyaMuonFitting}
\end{figure}

The mean number of photoelectrons ($N$) is estimated for those voltage settings, 
from the response to a single photoelectron calibration using the UV lamp.
In a first step, the electron-peak charge distribution of the UV lamp runs
is fit by Eq.~\ref{eq:Polyafunction} (where $N=1$),
in order to estimate the parameters $Q_e$ and $\theta$.
The electron-peak charge values $(Q_i)^n_{i = 1}$ of the 150~GeV muon run (where $n$ is the number of values)
were then used to define a likelihood function

\begin{equation}
\mathscr{L}(N|(Q_i)^n_{i = 1}) = \prod^n_{i = 1} \Bigg(\sum^\infty_{j = 0} \frac{N^j e^{-N}}{j!} \times A(Q_i|j,Q_e,\theta)\Bigg)
\label{eq:Likelihood} 
\end{equation}

where $Q_e$ and $\theta$ are results of previous fit,
$N$ is the mean number of photoelectrons per muon including the geometrical acceptance
and $A(Q)$ is the Polya-function defined in Eq.~\ref{eq:Polyafunction}.
This function was then maximized in order to estimate $N$.
An example of the results of these two fits is shown in Fig.~\ref{fig:PolyaMuonFitting},
while the value obtained is $N=10.4 \pm 0.4$.
The uncertainty of this estimation is dominated by the fit uncertainty of the UV lamp runs.

\section{Conclusions}
\label{sec:Conclusions}
In this paper, we present a new detector concept, called PICOSEC, composed of a ``two-stage'' Micromegas detector
coupled to a Cherenkov radiator and equipped with a photocathode.
The good timing resolution performance for single photoelectrons ($\sigma_t\sim 76$\,ps)
and for 150\,GeV muons ($\sigma_t\sim 24$\,ps) is promising and motivates further development towards practical applications.
Among the significant issues to be addressed to ensure suitability for a large area detector to be used in high rate experiment
are: 1) the development of efficient and robust photocathodes (or secondary emitters),
and 2) scalability, including the development of the corresponding readout electronics.

\section*{Acknowledgements}
We acknowledge the support of the RD51 collaboration, in the framework of RD51 common projects.
We also thank K. Kordas for valuable suggestions concerning the analysis of the data.
J.~Bortfeldt acknowledges the support from the COFUND-FP-CERN-2014 program (grant number 665779).
M.~Gallinaro acknowledges the support from the Funda\c{c}\~ao para a Ci\^{e}ncia e a Tecnologia (FCT), Portugal.
D.~Gonz\'alez-D\'iaz acknowledges the support from MINECO (Spain) under the Ramon y Cajal program (contract RYC-2015-18820).
F.J.~Iguaz acknowledges the support from the Enhanced Eurotalents program (PCOFUND-GA-2013-600382).
S. White acknowledges partial support through the US CMS program under DOE contract No. DE-AC02-07CH11359.

\bibliographystyle{JHEP}
\bibliography{20180222_Picosec_RevisionNIMA}
\end{document}